\documentclass[]{aa} 
\usepackage[usenames,dvipsnames]{xcolor}

\usepackage{graphicx}
\usepackage[export]{adjustbox}
\usepackage{txfonts}
\usepackage{multirow}

\usepackage{natbib,twoopt}
\usepackage[breaklinks=true]{hyperref} 
\usepackage{xspace}
\usepackage[normalem]{ulem}
\usepackage{multicol}

\newcommand{\vrms}{\ensuremath{\varv_{\textrm{rms}}}\xspace}

\newcommand{\Msol}{\ensuremath{\,\mathrm{M_\odot}}\xspace}
\newcommand{\Zsol}{\ensuremath{\,\mathrm{Z_\odot}}\xspace}

\newcommand{\Rsol}{\ensuremath{\,\mathrm{R_\odot}}\xspace}

\newcommand{\Lsol}{\ensuremath{\,\mathrm{L_\odot}}\xspace}

\newcommand{\cms}{\,cm\,s$^{-1}$\xspace}

\newcommand{\GG}[1]{}

 \def\simle{\mathrel{\hbox{\rlap{\hbox{\lower4pt\hbox{$\sim$}}}\hbox{$<$}}}}
 \def\simgr{\mathrel{\hbox{\rlap{\hbox{\lower4pt\hbox{$\sim$}}}\hbox{$>$}}}}

\graphicspath{%
 {./figs/}%
}

\begin{document}
   \title{Internal circulation in tidally locked massive binary stars - Consequences for double black hole formation }

   \author{B. Hastings     \inst{1}
            \and
            N. Langer \inst{1,2}
            \and
            G. Koenigsberger    \inst{3}
            }
    \institute{Argelander-Institut f\"{u}r Astronomie, Universit\"{a}t Bonn, Auf dem H\"{u}gel 71, 53121 Bonn, Germany \\
     \email{bhastings@astro.uni-bonn.de}\\
     \and     
    Max-Planck-Institut   f\"{u}r   Radioastronomie,   Auf   dem   H\"{u}gel   69, 53121 Bonn, Germany  \\
     \and
     Instituto de Ciencias F\'{i}sicas, Universidad Nacional Aut\'{o}noma de M\'{e}xico, Ave. Universidad S/N, Cuernavaca, Morelos 62210, M\'{e}xico
                }

    \authorrunning{Hastings, Langer and Koenigsberger}         
    \titlerunning{Internal circulation in tidally locked massive binary stars }

 
\abstract
   { Steady-state currents, so-called Eddington-Sweet circulation, result in the mixing of chemical elements in rotating stars, and in extreme cases lead to a
homogeneous composition. Such circulation currents are also predicted in tidally deformed binary stars, which are thought to be progenitors of double black-hole merger events.
} 
   {This work aims to quantitatively characterise the steady-state circulation currents in components of a tidally locked binary system and to explore the effects of such currents on numerical models. }
   {{Previous results describing the circulation velocity in a single rotating star and a tidally and rotationally distorted binary star are used to deduce a new prescription for the internal circulation in tidally locked binaries. We explore the effect of this prescription numerically with a detailed stellar evolution code for binary systems with initial orbital periods between 0.5 and 2.0 days, primary masses between 25 and 100 \Msol and initial mass-ratios $q_i$ =0.5,0.7,0.9,1.0 at metallicity Z=\Zsol /50 .}}
   {When comparing circulation velocities in the radial direction for the cases of a single rotating star and a binary star, it is found that the average circulation velocity in the binary star may be described as an enhancement to the circulation velocity in a single rotating star. This velocity enhancement is a simple function depending on the masses of the binary components and amounts to a factor of approximately two when the components have equal masses. After applying this enhancement to stellar models, it is found that the formation of double helium stars through efficient mixing occurs for systems with higher initial orbital periods, lower primary masses and lower mass ratios, compared to the standard circulation scenario. Taking into account appropriate distributions for primary mass, initial period and mass ratio, models with enhanced mixing predict 2.4 times more double helium stars being produced in the parameter space than models without.}
  {We conclude that the effects of companion-induced circulation have strong implications for the formation of close binary black holes through the chemically homogeneous evolution channel. Not only do the predicted detection rates increase but double black-hole systems with mass ratios as low as 0.8 may be formed when companion-induced circulation is taken into account.}

   \keywords{stars: massive --
             stars: rotation --
             stars: evolution
             }

\maketitle
%

\section{Introduction}

The existence of grand-scale circulations in stars was first proposed by \citet{1924MNRAS..84..665V}, who noticed that in a single rotating star the surfaces of thermal equilibrium and surfaces of hydrostatic equilibrium do not coincide, meaning that both thermal and hydrostatic equilibrium cannot coexist anywhere in the star. This results in circulation currents being driven inside the star. The first quantitative description of such currents was given by \citet{1929MNRAS..90...54E}, while a more rigorous model was proposed by \citet{1950MNRAS.110..548S}, and therefore such currents are referred to as Eddington-Sweet circulation. 

The fundamental cause of Eddington-Sweet circulation is that the effective gravity of a rotating single star becomes distorted due to a centrifugal force from a spherical pattern (as found in a non-rotating star) to an ovalised shape (where gravity is weaker at the equator). This distortion changes the thermal structure of the star through the Von Zeipel Theorem \citep{1924MNRAS..84..665V} and leads to the previously mentioned mismatch between thermal and hydrostatic equilibrium. 

It is also possible for a star's structure to be disturbed by a binary companion, therefore, in analogy to the single rotating case, a star in a binary system should also host internal circulations. The existence of circulations in binary stars stands on just as rigorous a mathematical footing as that for single rotating stars; however, in comparison to the rotating single star case, the characterisation of this binary circulation has received little attention, although significant efforts in this area have been made \citep{1981MNRAS.194..583S,1982ApJ...261..265T}. 

The effects of companion-induced circulation {are} most apparent in close binary systems, where the main effect will be a mixing of material between the stellar core and envelope, adding fresh fuel to the burning zones \citep{2009A&A...497..243D}. As shown by \citet{1987A&A...178..159M} and \citet{2006A&A...460..199Y}, such mixing can have dramatic effects on the evolution of stars. In extreme cases, the mixing is strong enough to prevent a chemical gradient developing within the star, causing homogeneous evolution where the star undergoes blue-wards evolution in the Hertzsprung-Russel diagram and the expansion of the stellar envelope is repressed. 

In the age of gravitational wave astronomy, there is renewed interest in binary-specific mixing mechanisms as it has been proposed by \citet{2016A&A...588A..50M} and \citet{2016MNRAS.458.2634M} that double black hole merger events can originate from close massive binary systems with very efficient internal mixing such that they undergo homogeneous evolution leading to double helium stars. Without internal mixing each star in the binary system would expand until one component surpasses the second Lagrangian point of the binary orbit (so-called L2 overflow), leading to a common envelope situation which will result in stellar merger for such close binary systems (Menon et al. in prep.). The models of \citet{2016A&A...588A..50M} and \citet{2016MNRAS.458.2634M} did not include the circulation induced by a binary companion, therefore the range of binary systems that undergo homogeneous evolution may be affected by the inclusion of these mechanisms and could have strong repercussions for double black-hole merger events. 

Evidence for internal mixing in binary systems exists for a handful of observed systems, the most convincing of which is VFTS352, a system of two 30\Msol stars in the Large Magellanic Cloud with an orbital period of 1.1 days \citep{2015ApJ...812..102A,2019ApJ...880..115A}. When comparing VFTS352 to rotating single star models, it is found that the best match to the observed effective temperatures and luminosities is given by models that are rotating significantly faster than the observed rotation rates in the system. Rotational mixing, which increases its efficiency with rotation rate, is thus judged to be stronger in the components of VFTS352 than in single star models, suggesting that mixing processes in massive close binary stars may be more efficient than those in rotating single stars. Other binary stars that show evidence of progression along a chemically homogeneous evolutionary track are R145 \citep{2017A&A...598A..85S} and HD 5980 \citep{2014AJ....148...62K}.
 
A significant proportion \citep{2012Sci...337..444S} of O-type stars are thought to be in fact binary stars, hence our understanding of massive stars requires a detailed description of the effects of binarity, with mixing processes being one of those effects. Internal mixing affects the luminosity and surface element abundances of stars, which are in turn used to calibrate stellar models and measure quantities such as mass and age by comparison to numerical models. Therefore a solid description of mixing processes is necessary for the detailed study of massive stars in general.  

This work aims to characterise quantitatively the steady-state circulation currents in a tidally locked component of a binary system and numerically investigate the effects of this circulation in massive short-period binary systems using a one-dimensional detailed stellar evolution code. To do this we shall utilise the mathematical results of \citet{1982ApJS...49..317T,1982ApJ...261..265T}. 

In Section \ref{sec:circmodel} we first introduce the mathematical models of internal circulation for a single rotating star and for a binary component and summarise the results of \citet{1982ApJS...49..317T,1982ApJ...261..265T}. A quantitative comparison between the two is made in Section \ref{sec:circ_comp} which allows us to model binary circulations in a one-dimensional stellar evolution code. We investigate the implications of our theoretical results in Section \ref{sec:MESAmodels} which outlines our numerical approach and Section \ref{sec:MESAresults} which presents the results of our simulations. Other mixing mechanisms that could feasibly operate and implications for double black hole merger events are discussed in Section \ref{sec:disc}. The final conclusions are give in Section \ref{sec:conc}.

\section{A mathematical model of circulation \label{sec:circmodel}}

The problem of circulation inside a star is a complex question in 3-dimensional fluid mechanics, therefore a number of simplifying assumptions must be made. We follow the mathematical formulation of \citet{1982ApJ...261..265T} to describe the circulation in the radiative zone of a tidally and rotationally distorted star. Their assumptions are as follows 

\begin{enumerate}
\item The orbit of the system is circular with each star rotating about its own axis with a period equal to the period of the binary orbit,  {such a system is said to be synchronised or tidally-locked}
\item The polar axes of the binary components are parallel to each other and perpendicular to their orbital velocity vector
\item Each star is acted on only by the tidal force of its companion
\item The system is detached
\item The companion of each star is modelled as a point mass
\item The stars are not significantly distorted away from sphericity
\item The stars rotate as solid-bodies 
\item The stars are on the zero-age-main-sequence such that there are no buoyancy forces arising from chemical gradients
\item The stars are non-magnetic
\item The circulation propagates in the radiative zone only
\item The radiative zone is inviscid
\end{enumerate}

We now discuss the justifications, if they exist, for each assumption. Many short-period binary systems are observed to be synchronised and to have nearly circular orbits. This is explained by dynamical tides that circularise the orbit and enforce spin-orbit coupling on timescales that are relatively short \citep{1975A&A....41..329Z,1977A&A....57..383Z}. For example the predicted synchronisation time of a binary star consisting of two 10\Msol stars with a period of 2 days is approximately $1\%$ of the nuclear timescale \citep{1975A&A....41..329Z}. For the same system the circularisation timescale is of the order $10\%$ of the nuclear timescale \citep{2002MNRAS.329..897H}. Thus massive short-period binary systems are expected to be circularised and synchronised, and assumption 1 is valid in many systems. 

For very luminous or extreme mass-ratio systems, a star may be affected by the radiation of its companion, which can drive a weak circulation that is unlikely to have any hydrodynamical effect or result in significant mixing \citep{1982ApJ...261..273T}, supporting assumption 3.

{In a contact binary, \citet{2009MNRAS.397..857S} envisioned that the circulation pattern must connect both stellar cores, and thus contact binaries require a different treatment. However the impact of circulation currents on the evolution of a star is mainly determined by the circulation strength at the core-envelope interface, as this is where fresh fuel is mixed into the core. In most cases the core-envelope interface would not be greatly affected by contact, thereby contact and non-contact systems are expected to behave similarly, giving justification for assumption 4. }

Assumptions 5 and 6 exist to simplify the problem and are only strictly justified for systems with orbital separations much larger than the stellar radii. 

Assumption 7 is valid owing to the action of the Tayler-Spruit dynamo \citep{2002A&A...381..923S}, which causes angular momentum to be transported throughout the stellar interior via magnetic interactions such that near solid-body rotation is achieved. The Tayler-Spruit dynamo is used to explain relatively slow rotation rates of white dwarves \citep{2008A&A...481L..87S} and young pulsars \citep{2005ApJ...626..350H}.

{A standard assumption is that stars enter the zero-age-main-sequence with a uniform chemical composition, meaning that no buoyancy forces act upon the circulation. Although during the star-formation process, material may become partially segregated, no process, such as nuclear burning, occurs in stars before the zero-age-main-sequence that can introduce significant chemical gradients. }

{Assumption 9 is a simplifying assumption only, although magnetic fields can introduce asymmetric distortions that may influence circulation currents \citep{1963MNRAS.126...67R,1977MNRAS.178...51M}. }

{Typical circulation velocities are not greater than and are mostly much slower than 10\cms (see Fig \ref{fig:Ukcomp}), while convective velocities are typically of the order $10^4$ \cms (see also diffusion coefficients plotted in Fig. \ref{fig:profiles}). Thus, the circulation currents are efficiently smoothed out by convection, and it is safe to ignore circulation in convective regions as in assumption 10.}

{Viscosity may be taken into account at the expense of complicating the final solutions, and given that \citet{1982ApJS...49..317T} find that their 'results imply that the circulation velocities do not depend much on the values of the coefficient of viscosity', we choose to work with the inviscid solutions. }

{The mathematical approach is to find the mean velocity of steady-state motion by expanding the Navier-Stokes equation about hydrostatic equilibrium to first order in the dimensionless parameter $\epsilon$, which characterises the strength of rotation and is given by}

\begin{align}
\epsilon = \frac{\Omega^2 R^3}{GM} = \bigg(\frac{\Omega}{\Omega_{crit}}\bigg)^2 ,
\end{align}
where $\Omega$ is the angular velocity of a star's rotation, $G$ the gravitational constant, $R,M$ the stellar radius and mass respectively and $\Omega_{crit}$ the critical angular velocity such that the star becomes gravitationally unbound at this angular velocity. It is thus clear that $\epsilon$ can never be greater than 1 and that $\epsilon$ is a measure of the strength of rotation. In a synchronised binary, Kepler's third law can be used to express the parameter $\epsilon$ as 

\begin{align}
\epsilon = \frac{G( M+M_{comp})}{d^3} , \label{Eq:kepler}
\end{align}
where $M_{comp}$ is the mass of the companion and $d$ the orbital separation.

The detailed derivation of circulation in a rotating single and binary star is given in \citet{1982ApJS...49..317T} and \citet{1982ApJ...261..265T}, respectively, upon which our treatment is entirely based.

\subsection{Circulation in a rotating single star \label{sec:ESC}}

{The potential of a barotrope \citep{1942ApJ....96..124K} is }
\begin{align}
V= A_2 \phi_2(r) - \frac{1}{3}\omega_0^2 r^2,
\end{align}
{where $A_2$ is a constant which ensures that there is a continuous potential inside and outside of the star, defined as}
\begin{align}
& A_2 = \frac{1}{3} \omega_0^2 \frac{5R^2}{3\phi_2(R) + R\phi'_2(R)}, \label{Eq:A2} \\
& \omega_0^2 = \frac{GM}{R^3}.
\end{align}
The function $\phi_2(r)$  satisfies the differential equation 
\begin{align}
\phi_k '' + \frac{2}{r} \phi_k' + \phi_k( 4\pi G \frac{\rho \rho '}{p'} - \frac{k(k+1)}{r^2}) =0 ,\label{Eq:phik}
\end{align}
with $k=2$, $\rho$ being the mass density and the boundary conditions 
\begin{align}
\phi_k(r=0) = \phi_k '(r=0) =0 \label{Eq:phikBoundary}.
\end{align}
A dash denotes a derivative with respect to the radial co-ordinate ( $p' \equiv \frac{dp}{dr}$ ).

By employing the expressions above, expansion about hydrostatic equilibrium gives the components (to 1st order in $\epsilon$) of the circulation velocity in the radial and colatitudinal directions as 
\begin{align}
u_r (r, \mu) &= \epsilon U_S(r) P_2(\mu) \label{Eq:Ur} , \\
u_{\mu} (r, \mu) &= \epsilon V(r) (1-\mu^2) \frac{dP_2(\mu)}{d\mu} \label{Eq:Umu} , 
\end{align}
where $\mu = cos(\theta)$ and $\theta$ is the angle from the star's pole (colatitude), $P_2$ is the second order Legendre polynomial, $U_S(r)$ and $V(r)$ are velocity magnitudes in the radial and colatitudinal directions respectively. The velocity magnitudes $U_S(r)$ and $V(r)$ are related to each other through the continuity equation, giving
\begin{align}
V(r) = \frac{1}{6 \rho r^2} \frac{d}{dr}(\rho r^2 U_S(r)) . 
\end{align}

The most important quantity for internal mixing is the circulation velocity magnitude in the radial direction, given by 
\begin{align}
U_S(r) &= \frac{2Lr^4}{G^2m^3} \frac{n+1}{n-1.5} \bigg[h' + \left(\frac{2}{r}- \frac{m'}{m}\right)h\bigg] \label{Eq:US} , 
\end{align}
where 
\begin{align}
& h(r) = A_2 \phi_2(r).\label{Eq:h}
\end{align}
In these expressions $L$ is the total stellar luminosity, $m$ the Lagrangian mass co-ordinate, $r$ the Eulerian radial co-ordinate and $n$ is the effective polytropic index throughout the model, which is the value for which the pressure and density profiles fit the polytropic relation at a given point in the star.

\begin{figure} 
	\includegraphics[width=0.95\linewidth]{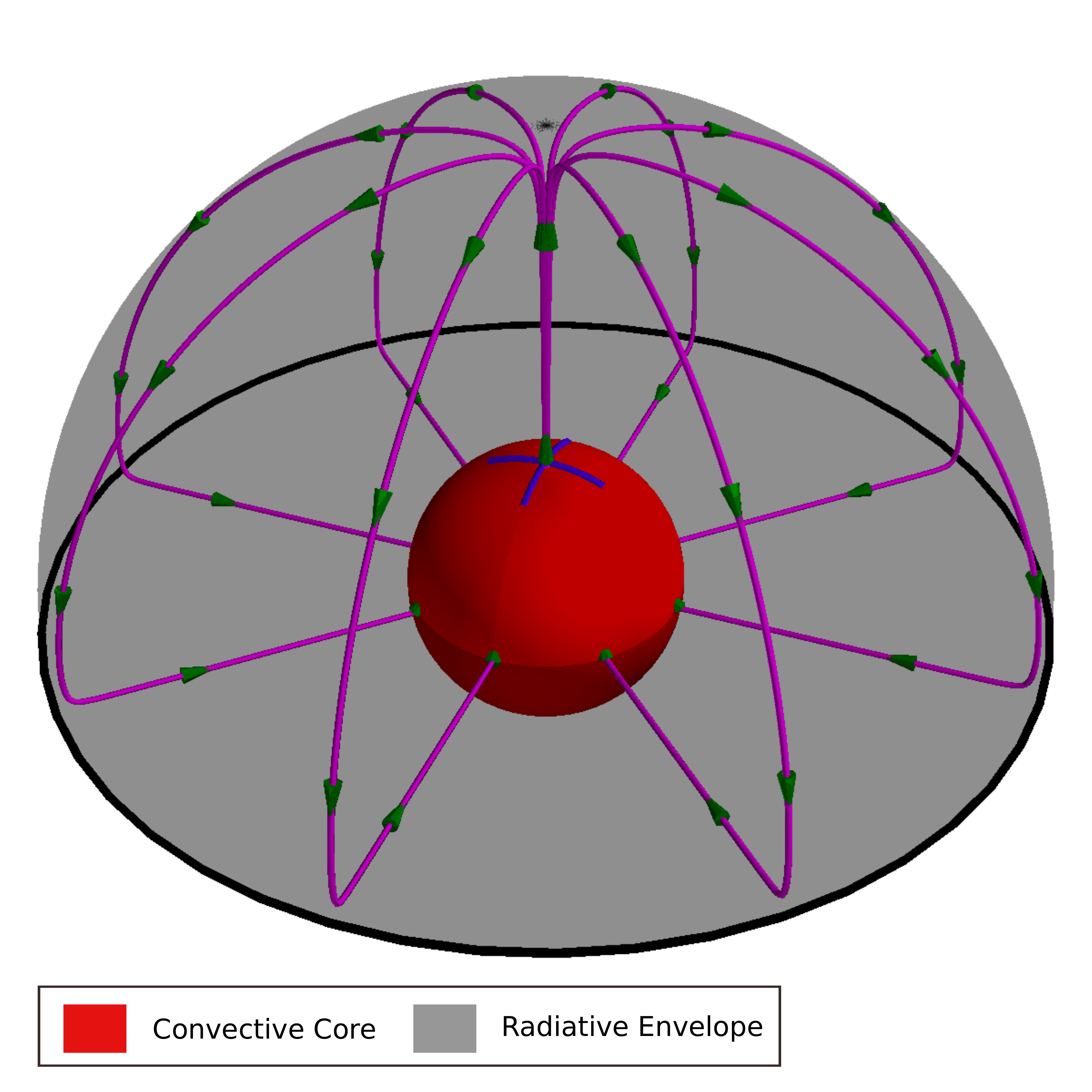}
	\centering
	\caption{Circulation pattern of Eddington-Sweet circulation in a single star Cowling point source model with electron scattering opacity, $M=$ 3\Msol, $R=$ 1.75\Rsol $L=$ 93\Lsol and $\epsilon=0.1$. Only the outer circulation cells in the northern hemisphere are shown. The green arrows give the direction of the motion. The central convective region is marked in red, the pole is marked with a blue cross and the radiative envelope is shown in opaque grey. This figure is a reproduction of Fig. 3 of \citet{1982ApJS...49..317T}}
	\label{fig:ESpattern} 
\end{figure}

We will now use Eqs. \ref{Eq:Ur} and \ref{Eq:Umu} and the velocity magnitudes for a Cowling point source model, as tabulated in \cite{1982ApJS...49..317T}, to show the 3-dimensional circulation pattern for a single rotating star, in Fig. \ref{fig:ESpattern}. This star has electron scattering opacity, mass of 3\Msol and rotation parameter of $\epsilon=0.1$. For simplicity only the outermost circulation cells are shown, which define the radius of the star. The other circulation cells are nested within these and have almost the same morphology. It is seen that the currents rise at the pole and retreat back into the star at the equator. The circulation is symmetric about the equatorial plane. It is remarkable that this pattern is very ordered and simple, with currents fixed in a meridional plane. The circulation pattern is largely a product of the symmetry of the system.


\subsection{Circulation in a binary component \label{sec:binaryC}}

We will now give an overview of the solutions to the problem of a star which is both tidally and rotationally distorted as given by \citet{1982ApJS...49..317T}. A binary system is described in standard spherical coordinates where the origin is the centre of mass of the star in question, the $x$-axis points towards the centre of gravity of the companion and the $z$-axis points towards the pole of the star in question. In these coordinates the plane described by $ \varphi=0$ is the sub-binary meridian. {The mathematical treatment of the binary case is identical to that of single star case, with of course a different gravitational potential.} The potential, $W$, of a rotating star with a binary companion is then 
\begin{align}
W = \frac{1}{3} \Omega^2 r^2 [1-P_2(\mu)] + \frac{GM_{comp}}{d} \sum_{k=2}^{4}  \frac{r^k}{d^k} P_k(\nu) \label{Eq:potential} ,
\end{align}
as found by \citet{1933MNRAS..93..449C} where $\mu= cos(\theta)$ , $\nu=sin(\theta)cos(\varphi)$ and $P_k$ represents the Legendre polynomial of order $k$. {We note that for the unique case of point-mass companion, Eq. \ref{Eq:potential} is exact. In all other cases, the potential would contain an infinite summation.}

The potential $W$ is a combination of 4 different potentials, one arising from the rotation of the star (the first term) and 3 arising from the binary companion (the second term). The overall 3-dimensional velocity field will then be a superposition of terms originating from each one of these potentials. This field is 
\begin{align}
u_{r}(r, \nu, \mu) &=\epsilon\bigg[ U_S(r) P_2(\mu) + \sum_{k=2}^{4}  U_k(r) P_k(\nu) \bigg] \label{Eq:binaryR},\\ 
u_{\mu}(r, \nu, \mu) & = \epsilon\bigg[ V(r) (1-\mu^2) \frac{dP_2(\mu)}{d\mu}  + \sum_{k=2}^{4}  V_k(r) (1-\mu^2) \frac{\partial P_k(\nu)}{\partial \mu} \bigg]  \label{Eq:binaryMU}, \\ 
u_{\varphi}(r, \nu, \mu) &= \epsilon \bigg[ \sum_{k=2}^{4} \frac{r V_k(r)}{\sqrt{(1-\mu^2)}} \frac{\partial P_k(\nu)}{\partial \varphi}  \bigg] . \label{Eq:binaryPHI}
\end{align}
The velocity $V_k(r)$ is related to $U_k(r)$ {by employing the continuity equation to give}
\begin{align}
V_k = \frac{1}{k(k+1)} \frac{1}{\rho r^2} \frac{d}{dr} ( \rho r^2 U_k)  . 
\end{align}

As in the single star case, each contribution to this velocity field consists of a dimensionless geometric term and a velocity magnitude term. The first terms of Eq. \ref{Eq:binaryR} and \ref{Eq:binaryMU} are exactly those of Eq \ref{Eq:Ur} and Eq \ref{Eq:Umu} and represent the circulation arising from the star's rotation. The terms in the summations in the equations above are due to the potential of the binary companion. Because the potential of a binary star can be decomposed into a summation of 3 terms (see Eq. \ref{Eq:potential}), we also get 3 different contributions to the velocity field, each with its own inviscid circulation velocity magnitude, $U_k(r)$, which is defined by the following equations 

\begin{align}
& U_k(r) = \frac{2Lr^4}{G^2m^3} \frac{n+1}{n-1.5} \bigg[h_k' + \left(\frac{2}{r}- \frac{m'}{m}\right)h_k\bigg] \label{Eq:USbinary} , \\ 
& h_k(r) = c_k \phi_k(r) , \\
& c_k = - \omega_0^2 \frac{M_{comp}}{M+M_{comp}} \left( \frac{R}{d}\right)^{(k-2)} \frac{(2k+1)R^2}{(k+1)\phi_k(R) + R\phi'_k(R)}  .\label{Eq:ckbinary}
\end{align}
Again $\phi_k$ can be found by solving Eq. \ref{Eq:phik} {and $c_k$ assures a smooth potential between the stellar interior and exterior.}

{Eqs. \ref{Eq:binaryR} - \ref{Eq:binaryPHI} and \ref{Eq:USbinary} - \ref{Eq:ckbinary} give the full description, inline with the initial assumptions, of circulation currents in a tidally-locked binary component. The dependence on the orbital separation is expressed in Eq. \ref{Eq:ckbinary} through the $\left( \frac{R}{d}\right)^{(k-2)}$ term and also in Eqs. \ref{Eq:binaryR} - \ref{Eq:binaryPHI} through the $\epsilon$ parameter, describing the strength of rotation. In a tidally-locked system, rotation is related to the orbital separation by Kepler's Third Law, Eq. \ref{Eq:kepler}. Therefore even when $k=2$ the dependence on the orbital separation is retained.}

\begin{figure*} 
	\includegraphics[width=0.99\linewidth]{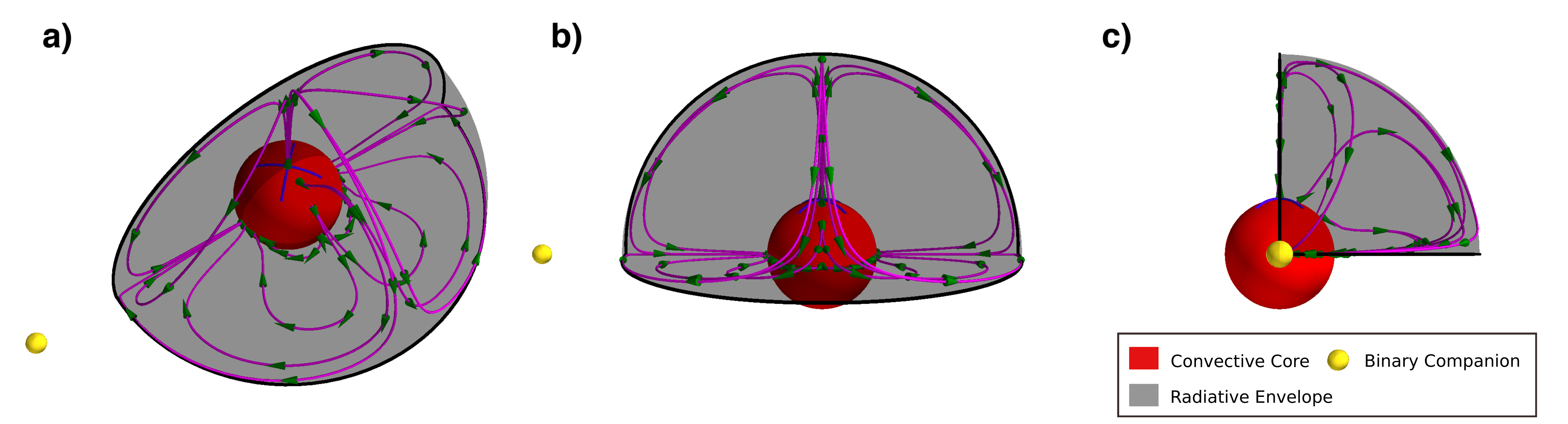}
	\centering
	\caption{Circulation pattern of a rotationally and tidally distorted Cowling point source model with electron scattering opacity, $M=$ 3\Msol, $M_{\textrm{comp}}=$3\Msol, $R=$ 1.75\Rsol $L=$ 93\Lsol and $\epsilon=0.1$. Three different perspectives are shown. The $k=3,4$ contributions to the velocity field due to the companion have been omitted. The companion is shown as a yellow sphere at an arbitrary orbital separation. The green arrows give the direction of the motion. The central convective region is marked in red, the pole is marked by a blue cross and the radiative envelope is shown in opaque grey.  }
	\label{fig:Binarypattern} 
\end{figure*}

Plotted in Fig. \ref{fig:Binarypattern} is this 3-dimensional velocity field for a binary component again with electron scattering opacity, $M=$ 3\Msol, $\epsilon=0.1$ and assuming a companion mass of $M_{\textrm{comp}}=$3\Msol. We ignore the $k=3,4$ terms, as they contribute little to the overall circulation pattern (see \citet{1982ApJ...261..265T} and Appendix \ref{sec:Ukcomparison}). Fig. \ref{fig:Binarypattern} shows that this is a very different pattern to the single rotating star shown in Fig. \ref{fig:ESpattern}. The circulation pattern is symmetric about the equatorial plane and about the plane passing through the poles and sub-binary meridian {whereas the Coriolis force could potentially change this, it does not enter into the linear order theory presented here.} As our goal is to model the circulation in a binary star using a one-dimensional stellar evolution code, we must make some kind of measurement of these two patterns to compare them.


\subsection{Limitations of the mathematical model}
By far the most constricting assumption in the mathematical description of internal circulation is that the star is not significantly distorted away from sphericity. As binary stars must have small orbital separations to become tidally locked early enough during their main-sequence evolution, the departures from spherical symmetry may be non-negligible. However, for our example system that is discussed in Sec. \ref{sec:exampleSystem}, the tidal deformation is smaller than 3\% and 5\% for the primary and secondary, respectively, based on the deformation parameter given in  \citet{2008EAS....29...67Z}. 

The mathematical model also relies on the Roche approximation, where the radii of the components are much smaller than the orbital separation so that the companion may be treated as a point mass. Our example system (Sec. \ref{sec:exampleSystem}) exhibits ratios of stellar radius to orbital separation of approximately 0.35 and 0.30 for the primary and secondary star respectively, so the Roche approximation may break down for massive tidally locked binaries.

{The solutions presented here are theoretical constructs and despite {the fact that} the assumptions are mostly justified, many phenomena exist which could alter the flow-patterns in Figs. \ref{fig:ESpattern} and \ref{fig:Binarypattern}. One of the most likely is the baroclinic instability which can set up wave-like azimuthal motions (see Appendix B of \citet{1982ApJS...49..317T}). Other processes include horizontal turbulence \citep{1992A&A...253..173C} and shear mixing which may only be investigated by a more sophisticated mathematical procedure. }

\section{A comparison of circulation patterns \label{sec:circ_comp}}

For convenience Table \ref{tab:VelComp} gives a summary of the different functions defining the circulation velocity in the radial direction for the single star and binary star cases.

\begin{table*}[h]
\centering
\caption{Summary of the functions defining the circulation in the radial direction, $u_r$, for the single and binary star cases}
\begin{tabular}{lll}
\hline
\hline
& Single Star                                                                  & Binary Star                                                                                                           \\ \hline
$u_r=$                                                                          & $\epsilon U_S(r) P_2(\mu)$                                                   & $\epsilon\big[ U_S(r) P_2(\mu) + \sum_{k=2}^{4}  U_k(r) P_k(\nu) \big]$                                               \\ 
\multirow{3}{*}{\begin{tabular}[c]{@{}l@{}}velocity\\  magnitude\end{tabular}} & $U_S =\frac{2Lr^4}{G^2m^3} \frac{n+1}{n-1.5} [h' + (\frac{2}{r}- \frac{m'}{m})h]$ & $U_k=\frac{2Lr^4}{G^2m^3} \frac{n+1}{n-1.5} [h_k' + (\frac{2}{r}- \frac{m'}{m})h_k]$                                      \\ 
                                                                               & $h(r) = A_2 \phi_2(r)$                                                      & $h_k(r) = c_k \phi_k(r)$                                                                                             \\ 
& $A_2 = \frac{1}{3} \omega_0^2 \frac{5R^2}{3\phi_2(R) + R\phi'_2(R)}$         & $c_k = - \omega_0^2 \frac{M_{comp}}{M+M_{comp}} (\frac{R}{d})^{(k-2)} \frac{(2k+1)R^2}{(k+1)\phi_k(R) + R\phi'_k(R)}$ \\ \hline
\end{tabular}
\label{tab:VelComp}
\end{table*}

Even though the circulation patterns inside a star are not one-dimensional, the effect of horizontal turbulence smooths out any  latitudinal or longitudinal dependent chemical gradients so that the effective transport of chemicals can be considered a one-dimensional problem \citep{1992A&A...253..173C}. 

To analyse the circulation patterns in one dimension, the root-mean-squared (rms) velocity is calculated in the radial direction over the whole surface of the star. The rms velocity must be considered because the mean velocity in the radial direction for any closed circulation pattern is zero. The rms radial velocity, denoted \vrms is

\begin{align}
\vrms &=  \frac{\sqrt{\int_{\varphi =0}^{\varphi=2\pi} \int_{\theta =0}^{\theta=\pi} u_r^2 d\theta d\varphi }}{  \sqrt{\int_{\varphi =0}^{\varphi=2\pi} \int_{\theta =0}^{\theta=\pi}  d\theta d\varphi}} . \label{Eq:vrms_def}
\end{align}

\subsection{A rotating single star \label{single_comp}}
Using Eq. \ref{Eq:Ur} and \ref{Eq:vrms_def} to calculate \vrms gives 

\begin{align}
\vrms &=  \sqrt{ \epsilon ^2 U_S^2(r)} \sqrt{\frac{1}{2\pi^2}} \sqrt{\int_{\varphi =0}^{\varphi=2\pi} \int_{\theta =0}^{\theta=\pi} P_2(cos(\theta))^2 d\theta d\varphi }  ,  
\end{align}
which results in 
\begin{align}
\vrms &= \epsilon |U_S(r)|   \sqrt{\frac{11}{32} } \approx 0.59 \epsilon |u_S(r)| . \label{Eq:singleResult}
\end{align}
This result is used below to compare the circulation in a binary star to that in a single star.

\subsection{A binary star \label{binary_comp}}

The inviscid radial velocity magnitudes, $U_k$ and $U_S$ are defined by the solution to the differential equation Eq.\ref{Eq:phik}. It is not evident how these solutions vary with $k$ and therefore it is in general difficult to compare the functions $U_k$ and $U_S$. The exception is $U_2$ and $U_S$, which both use the $k=2$ solution to Eq. \ref{Eq:phik}. When both the single star and binary component do not suffer significant deformation (which is a starting assumption in Sec. \ref{sec:circmodel}), they have the same structure and we see that $U_2$ and $U_S$ differ only by the constants $c_2$ and $A_2$ (see Table \ref{tab:VelComp}), which are related to each other by 

\begin{align}
c_2 &= - 3 \frac{M_{comp}}{M+M_{comp}} A_2. 
\end{align}
In terms of $q_*$, the effective mass ratio, $q_* = M_{comp}/M$, this expression becomes
\begin{align}
c_2  &= - 3 \frac{q_*}{q_*+1} A_2 . \label{Eq:c2a2}
\end{align}
Note that $q_*$ will change depending on which member of the binary is being considered, so that $q_* >1$ for the least massive star and $q_* <1$ for the most massive star in the system. 

Using Eq. \ref{Eq:c2a2} we may write
\begin{align}
U_2 = - 3 \frac{q_*}{q_*+1} U_S = Q U_S \label{Eq:U2USrel} ,
\end{align}
where for convenience we have set $Q= - 3 \frac{q_*}{q_*+1}$.

Armed with this relation, two extreme cases shall be considered, where $U_2$ dominates the companion-induced velocity field such that $U_2 >> U_3 >> U_4$ and the case where all terms contribute equally such that $U_2 \approx U_3 \approx U_4$.

\subsubsection{Extreme case i) : $U_2 >> U_3 >> U_4$ }
When we neglect $U_3$ and $U_4$ and apply the relation expressed in Eq.\ref{Eq:U2USrel},  Eq. \ref{Eq:binaryR} becomes
\begin{align}
U_{r}(r, \nu, \mu) &= \epsilon U_S(r)[   Q P_2(\nu) + P_2(\mu)] .
\end{align}
To calculate \vrms we compute 

\begin{align}
 \vrms = \epsilon |U_S(r)| \sqrt{\frac{1}{2\pi^2}} \Bigg[\int_{\varphi =0}^{\varphi=2\pi} \int_{\theta =0}^{\theta=\pi}Q^2 \bigg( P_2(cos(\varphi)sin(\theta))^2 \notag \\ + 
 P_2(cos(\theta))^2 + 2Q P_2(cos(\varphi)sin(\theta)) P_2(cos(\theta)) \bigg) d\theta d\varphi \Bigg]^{\frac{1}{2}} ,
\end{align}
which has the analytic solution
\begin{align}
\vrms &= \epsilon |U_S(r)|  \Bigg[\frac{49}{256}Q^2 +\frac{11}{32} -Q \frac{11}{32}\Bigg]^{\frac{1}{2}} . \label{Eq:case1Result}
\end{align}

\subsubsection{Extreme case ii) : $U_2 \approx U_3 \approx U_4$ \label{sec:Ex2}}

In the limit that $U_2 \approx U_3 \approx U_4$, Eq.\ref{Eq:binaryR} becomes 
\begin{align}
U_{r}(r, \nu, \mu)&= \epsilon U_S(r) [ Q\sum_{k=2}^{4}   P_k(\nu) +  P_2(\mu)].
\end{align}
{The calculation is provided in Appendix \ref{sec:case2calc}, giving \vrms as }
\begin{align} 
\vrms & \approx \epsilon |u_S(r)|  \sqrt{0.3410 Q^2  +0.34375  -  0.2295 Q}  \, . \label{Eq:case2Result}
\end{align}

It is apparent that Eqs. \ref{Eq:singleResult}, \ref{Eq:case1Result} and \ref{Eq:case2Result} describing the circulation velocities in one dimension for a single rotating star and a binary component are of the same functional form, that is proportional to $\epsilon |U_S(r)|$ and a factor depending on the effective mass ratio, $q_*$. To compare how much circulation is enhanced in the binary case, we can divide Eqs. \ref{Eq:case1Result} and \ref{Eq:case2Result} by Eq.\ref{Eq:singleResult} to give the factor by which a binary companion increases the circulation velocity in a binary star compared to an equivalent single star with the same $\epsilon$ parameter. We shall name this the velocity enhancement factor, $f_{\textrm{ES}}$.

\begin{figure} 
	\includegraphics[width=0.95\linewidth]{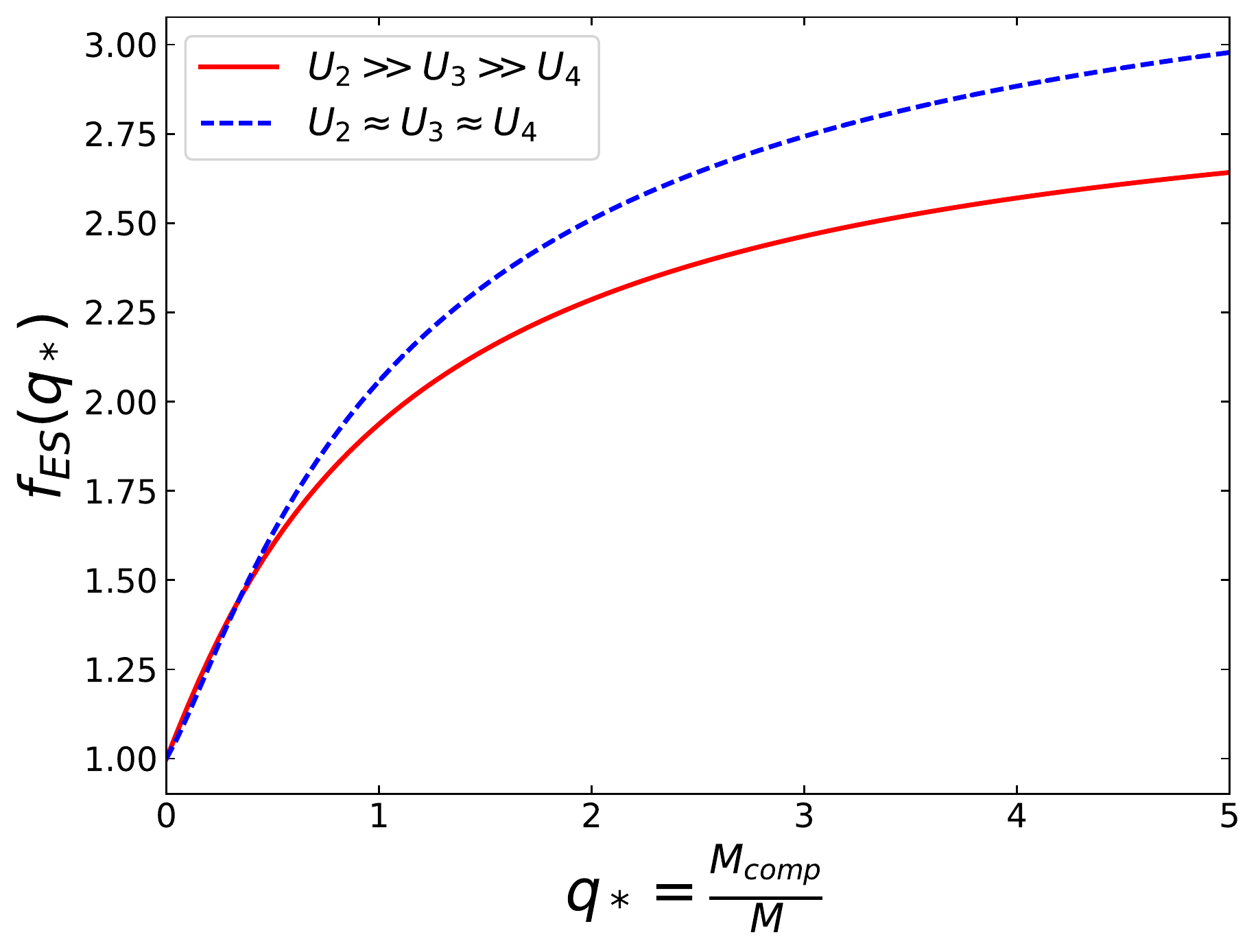}
	\centering
	\caption{Factor by which averaged Eddington-Sweet circulation velocity is enhanced in a tidally locked binary component as a function of the effective mass ratio, $q_*$ for two approximations as given in the legend.}
	\label{fig:Boostfunc} 
\end{figure}

Fig. \ref{fig:Boostfunc} shows this enhancement factor, which remarkably demonstrates that there is not much difference between the two extreme cases considered. For $q_* <1$ there is almost no difference between the two cases while for $q_* >1$ the difference amounts to approximately $10\%$. For the least massive star in the binary system, $q_* >1$ and for the most massive star $q_* <1$, so that circulation in the least massive star is enhanced more than it is in the most massive star. This follows simply from the fact that the most massive star imparts a larger gravitational disturbance on its companion than its companion does to it. 

In the limit that the companion mass, $M_{comp}$ tends to 0, Eqs. \ref{Eq:case1Result} and \ref{Eq:case2Result} revert to \ref{Eq:singleResult}, and as one would expect, the velocity enhancement factor becomes 1. It is curious to note that in the limit that the companion mass becomes infinite, $q_*$ tends to infinity but $Q= - 3 \frac{q_*}{q_*+1}$ tends to $-3$, hence the velocity enhancement factor has a limit of approximately $3.0$ and $3.4$ for our two extreme cases respectively.  

A decision must be made as to which approximation is the most suitable. Using polytropic models, one can show that in general the condition $U_2 >> U_3 >> U_4$ is the most realistic (see Appendix \ref{sec:Ukcomparison}). Such a result could also be inferred by studying the circulation patterns arising from $U_2$, $U_3$ and $U_4$ plotted by \citet{1982ApJ...261..265T}. Although inspection of Fig. \ref{fig:Boostfunc} leads us to conclude that both cases give such similar results so as to make the choice of approximation nearly inconsequential.

The final result of our analysis is that for a fixed rotation parameter $\epsilon$, in a tidally locked binary component the averaged circulation velocity, compared to a single rotating star is increased by a factor of 
\begin{align}
f_{\textrm{ES}}= \sqrt{\frac{441}{88}   \bigg( \frac{q_*}{q_*+1}\bigg) ^2 +1 +  \frac{3q_*}{q_*+1}} .
\end{align}

This expression is remarkable as it only depends on the masses of the stars in the binary, while the strength of the companion-induced circulation depends on the orbital separation and the component masses.
As Eddington-Sweet circulation depends on a star's rotation, for a full description of the circulation in a binary component one needs to know about the rotation rate, the orbital separation and the component masses. In a tidally locked binary, the orbital separation is set by the rotation rate, and so only the rotation rate and component masses are required. The rotation rate sets the strength of Eddington-Sweet circulation and gives us the orbital separation, therefore the velocity enhancement factor needs only to depend on the component masses. 


\section{Numerical method\label{sec:MESAmodels}}

To investigate the implications of companion-induced mixing as derived in Sec. \ref{sec:circ_comp}, version r10398 of the one-dimensional detailed stellar evolution code {MESA \citep{Paxton2011, Paxton2013, Paxton2015, Paxton2018, Paxton2019}} is used. The implemented physics is the same as that of \citet{2016A&A...588A..50M}, where massive binary systems without enhanced mixing were computed. However since that work was done, the treatment of angular momentum loss from binary systems was improved. With the improved version of MESA, the results presented here for our models without enhanced mixing will vary slightly from those of \citet{2016A&A...588A..50M}. 

{The opacities used in our calculations are given by the OPAL project \citep{Iglesias1993,
Iglesias1996}. For the nuclear reaction rates, we employ those of JINA REACLIB \citep{Cyburt2010}.}

Convection is modelled using standard mixing length theory \citep{1958ZA.....46..108B} with mixing length parameter $\alpha=1.5$ with the Ledoux criterion. We implement semi-convection as described by \citet{1983A&A...126..207L} with efficiency parameter $\alpha_{SC} =1.0$. Convective overshooting is modelled with a step function with $\alpha_{OV}=0.345$ as in \citet{2011A&A...530A.115B}.

The stellar wind prescription follows that outlined in \citet{2006A&A...460..199Y}. For stars with $Y_S <0.4$ mass-loss rates are given by the recipe of \citet{2001A&A...369..574V}. Where $Y_S >0.7$ , the prescription of \citet{1995A&A...299..151H} reduced by a factor of 10 is used. Where $0.4<Y_S<0.7$ , the mass-loss rate is determined by an interpolation between the two schemes. For both schemes a metallicity scaling of $(Z /Z_{\odot}) ^{0.85}$ was used. Rotationally enhanced mass-loss is handled by the method of \citet{2000ApJ...544.1016H} whereby the relationship between mass-loss with and without rotation is given by

\begin{align}
\dot{M} (\Omega) = \dot{M}(\Omega=0)  \Bigg[ \frac{1}{1- \frac{\Omega}{\Omega_ {\textrm{crit}}(\Gamma)}} \Bigg] ^{0.43} ,
\end{align}
 with the critical rotation rate being 
 
\begin{align}
\Omega_ {\textrm{crit}} (\Gamma) = \sqrt{\frac{GM}{R^3} (1-\Gamma)} \ ,
\end{align}
and $\Gamma $ the ratio of the stellar luminosity to the Eddington luminosity, $\Gamma = L/ L_{\textrm{Edd}}$. As the model approaches the critical rotation rate, mass-loss rates become divergent. To avoid this issue, once ${\Omega} /{\Omega_ {\textrm{crit}}}$ reaches 0.98, we switch to an implicit mass-loss rate, which is calculated such that the fraction ${\Omega} /{\Omega_ {\textrm{crit}}}$ remains 0.98.

Internal angular momentum transport is accomplished via the Tayler-Spruit dynamo \citep{2002A&A...381..923S}, which enforces near solid-body rotation through magnetic interactions. This is implemented numerically by the method of \citet{2005A&A...435..247P}. Tidal synchronisation is modelled according to \citet{2002MNRAS.329..897H} and \citet{2008A&A...484..831D}. {The criteria governing Roche-lobe overflow come from considering the Roche-lobe volumes and their volume-equivalent radii as per the prescription of \citet{Eggleton1983}, while mass-transfer rates are determined implicitly by demanding that the mass-donor stays inside its Roche-lobe. }

The mixing processes taken into account are Eddington-Sweet circulation, secular and dynamical shear instabilities and the GSF instability. The rotational mixing efficiency factor $f_c$ is set to $1/30$, as determined theoretically by \citet{1992A&A...253..173C} and the factor controlling the effects of chemical gradients on mixing $f_{\mu}$ is set to 0.1 as calibrated by \citet{2006A&A...460..199Y}. {The suppression of mixing by chemical gradients is included as a stabilising circulation with velocity proportional to the mean-molecular weight gradient and $f_{\mu}$ \citep{2000ApJ...544.1016H}.} Circulation is modelled as a diffusive process according to \citet{2000ApJ...544.1016H}, where the diffusion coefficient is calculated as the product of the circulation velocity and a typical length scale for the circulation. It is thus appropriate, as determined in Sec. \ref{binary_comp}, to  model the effects of companion-induced circulation by increasing the diffusion coefficients of Eddington-Sweet circulation by a factor of 
\begin{align}
f_{ES}= \sqrt{\frac{441}{88}   \bigg(\frac{q_*}{q_*+1}\bigg) ^2 +1 +  \frac{3q_*}{q_*+1}} , \label{Eq:modelboost}
\end{align}
with $q_* = M_{comp} / M$ being the effective mass ratio for the star in question. For the most massive star in the binary, $q_* <1$, while for the least massive star $q_* > 1$. The circulation velocity enhancement is achieved through the \texttt{D\_ES\_factor} control parameter in MESA. For models without companion-induced circulation we set $f_{ES}=1$.  

We compute grids of binary models with primary mass, $M_1$, ranging from log$(M_1 /M_{\odot})=1.4 $  to $2.0$ in intervals of 0.1 dex, such that the models are not massive enough to undergo the pair-instability supernova phenomenon. We choose initial orbital periods ranging from 0.5 to 2.0 days in intervals of 0.1 day. Four different initial mass-ratios are considered, $q_i = M_2/M_1=1.0 , 0.9, 0.7,0.5$ such that the mass of the secondary star, $M_2$ is the product of the primary mass and the mass-ratio. All models have an orbital eccentricity of 0 and rotation periods synchronised with the binary orbital period. 

The calculations are performed at a metallicity of $Z=Z_{\odot} /50$ where the initial helium abundance is the result of a linear interpolation in $Z$ between the primordial value $Y=$0.2477 at $Z=0$ \citep{2007ApJ...666..636P} and $Y=0.28$ at $Z=Z_{\odot}$ where we take $Z_{\odot}=0.017$ \citep{1996ASPC...99..117G}. {This relatively low metallicity is chosen because for higher metallicity systems, mass-loss through stellar winds causes a widening of the binary orbit which may result in double black-hole systems that are too wide to merge within a Hubble time (c.f Fig. 4 of \citet{2016A&A...588A..50M}). We note that as we only investigate the relative effect of enhanced circulation on the models, and that the minimum initial rotation rate needed for homogeneous evolution to occur is approximately constant with varying metallicity \citep{2006A&A...460..199Y}, we do not expect our analysis to be greatly dependent on the choice of metallicity.}

Termination occurs either at core helium depletion or when one component suffers L2 overflow (upon whence a merger is assumed to happen) or when the difference between central and surface helium mass fraction exceeds 0.2 ( meaning that chemically homogeneous evolution is not occurring, upon whence L2 overflow is assumed to eventually happen). When one component depletes helium, it is treated as a point mass until the other component depletes helium and the calculation is terminated. 

To determine the effects of the enhanced mixing in a binary system, we calculate two grids of models, without and with the companion-induced mixing described above. Models without enhanced mixing are named the Standard Grid, while those with enhanced mixing are named the Enhanced Grid. The inlists required to run the Standard Grid can be found in the binary test suite of MESA version r10398. 

\subsubsection{Uncertainties in numerical modelling}
An issue is the treatment of mass-loss, which is particularly important in a binary system as the binary orbit is influenced by both mass and angular-momentum loss through stellar winds. We are considering stars with initial masses up to 100\Msol, observations of which are rare and difficult to interpret. Therefore it is not certain that the mass-loss rates, as found by \citet{2001A&A...369..574V} and \citet{1995A&A...299..151H} for hydrogen rich and poor stars respectively, give an accurate picture of mass-loss in very massive stars. 

Furthermore the 3-dimensional mass loss pattern of a star in a close binary system is a complex affair. Due to the von Zeipel Theorem \citep{1924MNRAS..84..665V}, the temperature and flux across the stellar surface will be a function of both angular co-ordinates, and therefore so will the mass-loss rates. This may lead to a significant difference between effective mass-loss rates of single and binary stars. 

To model contact systems, the effects of energy transfer from one star to another are ignored, which for systems with initial mass ratios far from unity can play a role in the stellar evolution of both components. It is theoretically predicted and observed \citep{2004A&A...414..317K,2017A&A...607A..82M} that for contact systems with radiative envelopes, the difference in effective temperature between the stars in the system is very small, even when the mass-ratio is significantly different from unity. This suggests that energy is indeed transferred from one component to another during the evolution of a contact binary.

\section{Model results \label{sec:MESAresults}}
{In this section we compare the results of our model grid without enhanced mixing, the Standard Grid, and our model grid with enhanced mixing, the Enhanced Grid.}
\subsection{An example system \label{sec:exampleSystem}}
 To characterise the strength of mixing, the mixing timescale, $\tau _{\textrm{mix}}$ is calculated as \citep{1977A&A....56..211S}

\begin{align}
\tau_{\textrm{mix}} = \frac{4}{\pi^2} \Bigg[ \int _0 ^{R} \frac{dr}{\sqrt{D_{\textrm{tot}}}} \Bigg] ^{2},
\end{align}
where $D_{\textrm{tot}}$ is the sum of diffusion coefficients of all mixing processes operating within the star. The mixing timescale is interpreted as the time which is required for a particle to be mixed from the stellar centre to the photoshpere. When the fraction of the nuclear and mixing timescale, $\tau_{\textrm{nuc}} / \tau_{\textrm{mix}}$ is computed, the strength of internal mixing is easily assessed. The occurrence of homogeneous evolution is ruled by whether mixing processes can erode chemical gradients produced from nuclear burning at a fast enough rate such that chemical gradients do not inhibit mixing. Therefore chemically homogeneous evolution is expected to occur for values of $\tau_{\textrm{nuc}} / \tau_{\textrm{mix}} \gtrsim 1$. 

In the following we shall use the convention that the initially more massive star is named Star 1 and the initially less massive star is Star 2. These names will never be switched.

\begin{figure*} 
	\includegraphics[width=0.95\linewidth]{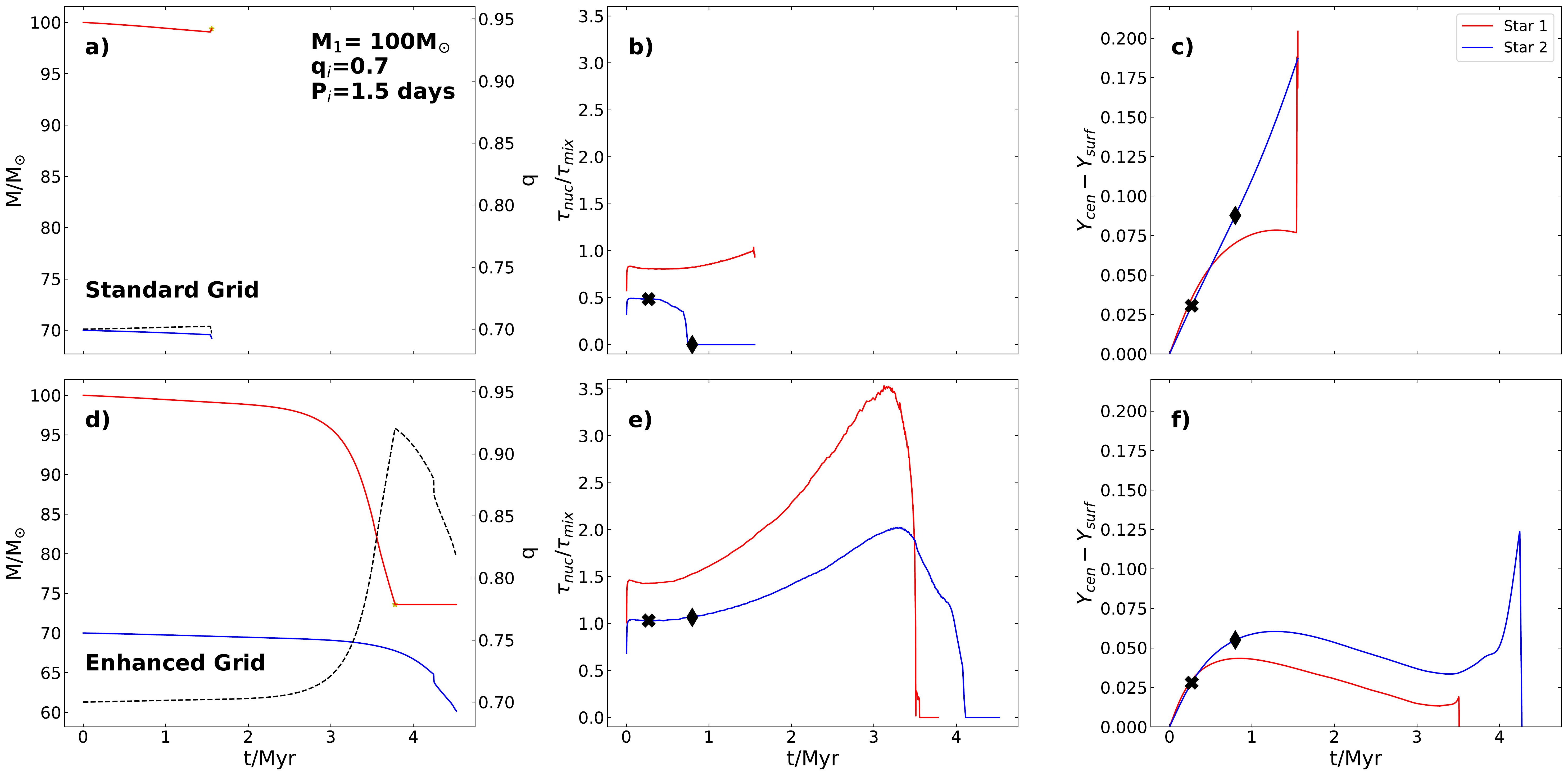}
	\centering
	\caption{Diagnostic plots for a system with primary mass, $M_1=$  100 \Msol, initial mass-ratio, $q_i=0.7$ and initial orbital period 1.5 days. The top panels represent the system in the Standard Grid (without enhanced mixing) and the bottom panels show the system in the Enhanced Grid (with enhanced mixing). The left panels show the evolution of each component's mass and the mass ratio, $q$ with a dotted line. The central panels show the ratios of nuclear and mixing timescales as a function of time. The right panels show the difference between central and surface helium mass fraction. Star 1 is plotted in red, Star 2 in blue. The black cross and diamond represent the times for which profiles of Star 2 are plotted in Fig.\ref{fig:profiles}. }
	\label{fig:Example} 
\end{figure*}

To understand the effects of our companion-induced circulation model we shall examine in detail a system with a primary mass, $M_1=100$ \Msol, initial mass ratio $q_i=0.7$ and initial period 1.5 days. For this system in the Standard Grid, Star 2 does not evolve homogeneously, whereas in the Enhanced Grid a double helium star is formed. Fig. \ref{fig:Example} shows the evolution of component masses, the ratios of nuclear and mixing timescales,  $\tau_{\textrm{nuc}} / \tau_{\textrm{mix}}$ and the difference between central and surface helium mass fraction for this example system in both Standard and Enhanced grids. 

{We first consider Star 2 at early times.  Figure \ref{fig:Example} b) shows that in the Standard Grid Star 2 is unlikely to undergo homogeneous evolution as the mixing timescale is around double the nuclear timescale. In the Enhanced Grid however, the diffusion coefficients of Eddington-Sweet circulation are increased by a factor of 2.1, which decreases the mixing timescale such that it is approximately equal to the nuclear timescale as shown in Fig. \ref{fig:Example} e).}

\begin{figure} 
	\includegraphics[width=0.99\linewidth]{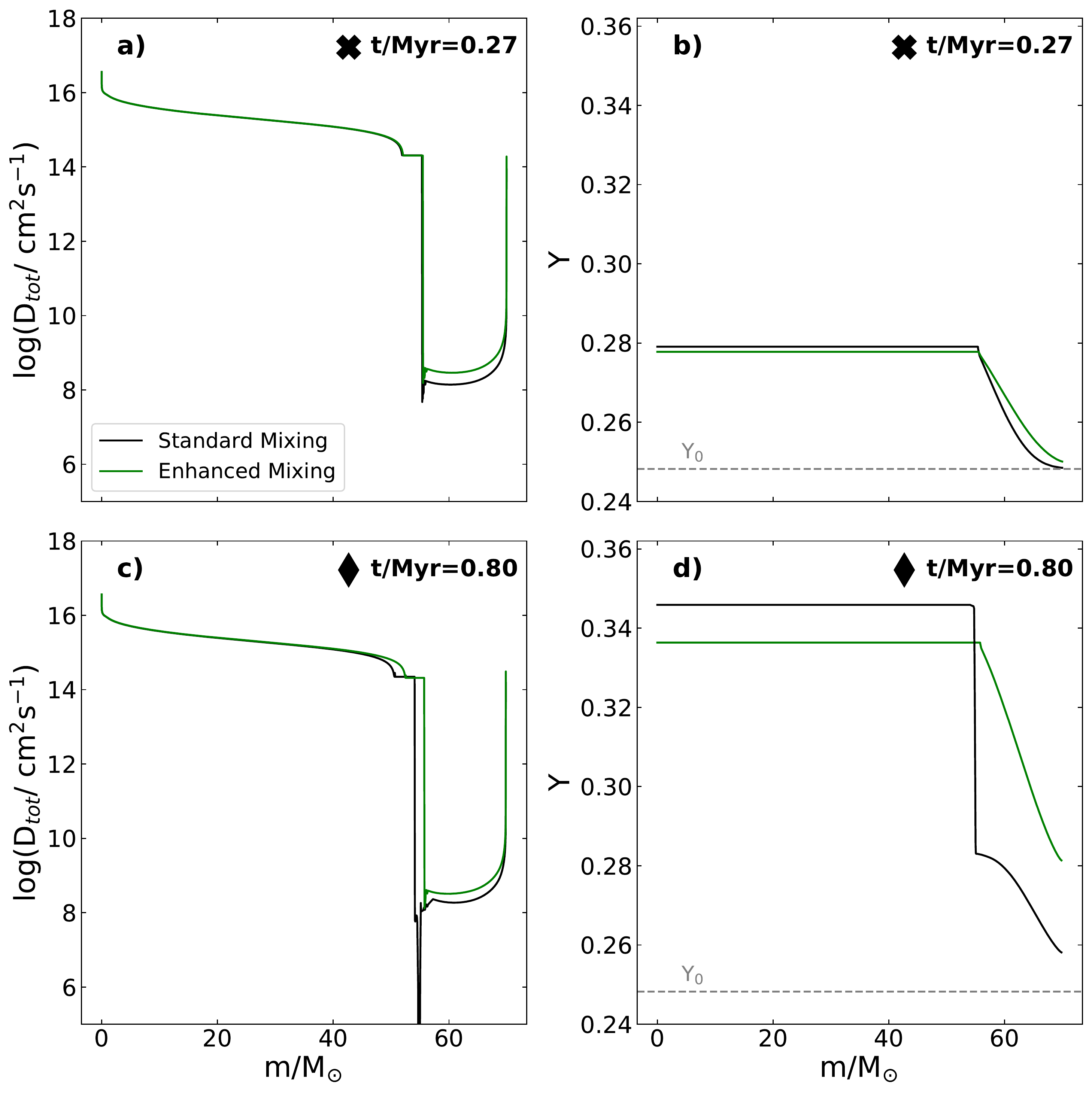}
	\centering
	\caption{ Internal profiles of Star 2 in the example system ($M_1=$  100 \Msol, $q_i=0.7$, $P_i=$1.5 days) showing the total diffusion coefficient, $D_{tot}$ (left panels), and helium mass fraction, Y (right panels), as a function of lagrangian-mass-coordinate. The star in the Standard Grid is marked in green and that in the Enhanced Grid in black. The top panels show profiles at an age of 0.27Myr, as marked by a cross in Fig.\ref{fig:Example}, while the bottom panels represent an age of 0.80Myr, as marked by a diamond in Fig.\ref{fig:Example}. The grey dashed line in the right panels represents the initial helium mass fraction, Y$_0$.}
	\label{fig:profiles} 
\end{figure}

{Looking at the chemical diffusion coefficient throughout the star in Fig. \ref{fig:profiles} a), we see that at early times, the Enhanced and Standard Grid models differ only by the increased diffusion coefficients of circulation in the envelope. Here we can easily distinguish between the convective core, where the diffusion coefficients are very high, and the radiative envelope where circulation operates with lower mixing efficiency. Similarly in Fig. \ref{fig:profiles} b), showing the helium mass fraction profiles at an age of 0.27Myr, not enough time has elapsed to allow either helium to be mixed to the surface, nor a strong chemical gradient to develop. In the model with enhanced mixing, even though only a negligible amount of helium has been mixed to the surface, the helium gradient between the core-envelope boundary and surface is shallower than in the Standard Grid model,  as shown in \ref{fig:profiles} b).}

{For the Standard Grid model this steeper helium gradient causes a slowing of the internal mixing through buoyancy forces, as evidenced by decreasing values of $\tau_{\textrm{nuc}} / \tau_{\textrm{mix}}$ in Fig. \ref{fig:Example} b). Looking at the internal profiles at an age of 0.8Myr in Fig. \ref{fig:Example} d), we see that when circulation is enhanced, appreciable amounts of helium have been mixed to the surface. In the standard mixing regime, a relatively small amount of helium has been mixed to the surface, however we see that a large step in helium abundance has developed at the core-envelope boundary, which provides a solid barrier to mixing. The effects of this barrier can be seen in Fig. \ref{fig:profiles} c) which shows very low diffusion coefficients at the core-envelope boundary. After this step in the chemical gradient has developed, mixing is shut off completely and the star evolves along the standard track, expanding until it interacts with its companion, causing the companion's surface helium abundance to increase and causing the calculation to terminate, as evidenced by Fig. \ref{fig:Example} c). }

During homogeneous evolution Fig. \ref{fig:Example} e) shows that  $\tau_{\textrm{nuc}} / \tau_{\textrm{mix}}$ evolves to larger values, meaning that mixing becomes more efficient. This occurs because a star undergoing homogeneous evolution evolves to hotter temperatures and approaches the Eddington luminosity \citep{1987A&A...178..159M}, which increases the light to mass ratio and thus causes the convective core to grow, this growth may be seen in panel c) of Fig. \ref{fig:profiles} by comparing the enhanced model to the standard model. As the diffusion coefficients associated with convection are some seven orders of magnitude larger than those describing Eddington-Sweet circulation, the mixing timescale naturally decreases when the extent of the convective core increases. This results in the stars evolving to larger values of  $\tau_{\textrm{nuc}} / \tau_{\textrm{mix}}$ in Fig. \ref{fig:Example} e).

Our example system in the Enhanced Grid shows the typical behaviour of a system that forms a double helium star without either component ever filling its Roche-lobe. Despite the lack of mass-transfer, the mass-ratio of the system evolves towards unity as the more massive star experiences higher mass-loss rates, as shown in Fig. \ref{fig:Example} d), giving a final mass-ratio somewhat closer to unity than the initial mass-ratio. However, owing to the fact that mass-loss from stellar winds is strongly metallicity dependant, very low metallicity systems will remain closer to their initial mass-ratios.


\subsection{Fates of systems}
\begin{figure} 
	\includegraphics[width=0.95\linewidth]{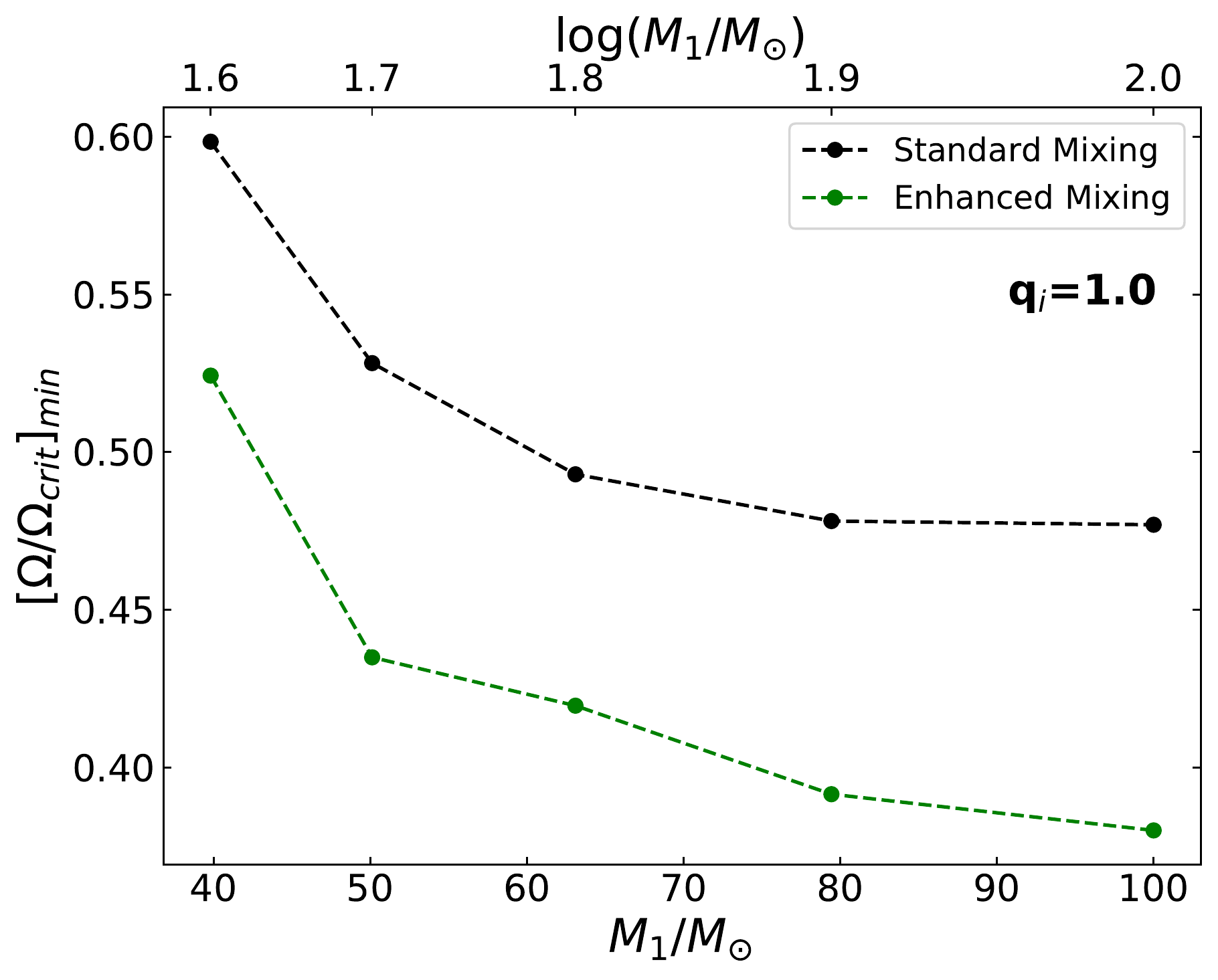}
	\centering
	\caption{Minimum initial fraction of critical angular velocity required for homogeneous evolution to occur as a function of primary mass for systems with $q_i=1.0$. Models from the Standard Grid are marked in black and those from the Enhanced grid in green.  }
	\label{fig:minrot} 
\end{figure}

{The major effect of our enhanced mixing prescription is to reduce the minimum rotation rate required for homogeneous evolution to occur, as shown in Fig. \ref{fig:minrot}, where the minimum initial rotation rate for  homogeneously evolving models is plotted as a function of mass for equal-mass systems. This plot shows stars with masses from log$(M_1/M_{\odot}) =1.6$ because $q_i=1.0$ systems with primary masses less than this value did not evolve homogeneously (see left panels of Fig. \ref{fig:gridResults}). For these systems the circulation velocity is enhanced by approximately a factor of two in each component, which causes homogeneous evolution to occur at fractions of critical angular velocity some 10-20 \% lower, depending on the mass, than for systems without enhanced mixing.}

{As stellar mass increases, the contribution of radiation pressure begins to dominate over gas pressure and the dimensionless adiabatic temperature gradient ($\nabla _{ad} = \left( \frac{d log T}{d log P} \right)_{ad} $) decreases from 0.4, the value for an ideal gas to 0.25, that of a radiative gas (see Ch. 13.2 of \citet{2012sse..book.....K}). The dependence on the subadiabaticity ($\nabla_{ad} - \nabla$) in many formulations of Eddington-Sweet circulation \citep{1974IAUS...66...20K,1983apum.conf..253Z,1987A&A...178..159M}, including that of the models presented here \footnote{MESA is based on the numerical treatment of \citet{2000ApJ...544.1016H} which, in turn uses the formalism of \citet{1974IAUS...66...20K} }, indicates that the stability of a fluid is described by the deviation from adiabaticity, with more stable systems having larger values of $\nabla_{ad} - \nabla$. Increased radiation pressure serves to reduce the dimensionless adiabatic temperature gradient, reducing the stability of the fluid and thus leading to circulation currents that propagate more easily than in the absence of radiation pressure. Thus, as the ratio of radiation pressure to gas pressure increases with stellar mass, more massive stars are expected to have more efficient circulation currents, hence resulting in lower rotation rates required to achieve homogeneous evolution, as seen in Fig. \ref{fig:minrot}. }

We identify three consequential effects of our enhanced mixing

\begin{enumerate}[i)]
  \item that stars of lower mass are able to evolve homogeneously
  \item that homogeneous evolution occurs in systems with longer initial periods
  \item fewer systems undergo mass-transfer events and contact phases, owing to larger initial orbital separations resulting from ii)
\end{enumerate}

{When the minimum rotation rate required for homogeneous evolution becomes low enough to be attained by a tidally-locked binary without suffering L2 overflow at the zero-age-main-sequence, stars of lower mass can evolve homogeneously in the enhanced mixing scenario. }

{The lowering of the minimum rotation rate required for homogeneous evolution means that systems with longer periods can undergo strong mixing and form double helium stars in comparison to the standard mixing scenario, as in tidally-locked systems the orbital period determines the stellar rotation. A side effect of this is that with enhanced mixing, binary orbits are wider and the stars are able to expand more without interacting with their companion, meaning that mass-transfer episodes become rarer, as discussed in detail in Sec. \ref{sec:mt} }

{The three effects of our enhanced mixing discussed above are visible in} Fig. \ref{fig:gridResults}, which shows the results of our binary evolution calculations. As homogeneous evolution did not occur in any model with $q_i = 0.5$, these models are subsequently ignored. We see that enhanced mixing results in systems with larger initial periods evolving homogeneously, causes systems with lower primary mass to evolve homogeneously, and causes fewer systems to evolve into a contact phase. We note that for the Enhanced Grid models with $q_i = 1.0$ and  log$_{10}(M_1/M_{\odot})=2.0$, we have not found the upper initial orbital period limit for homogeneous evolution to occur, at this initial mass-ratio, we strictly consider a lower limit.

In the Standard Grid, we see that no system with initial period greater than 1.6 days produces a double helium star. In comparison our Enhanced Grid models are able to evolve homogeneously at initial periods as high as 2.0 days for systems with initial mass-ratio  $q_i = 1$, 1.9 days for  $q_i = 0.9$ and 1.5 days for  $q_i = 0.7$. At every initial mass-ratio, for initial primary masses greater than log$_{10}(M_1/M_{\odot}) \approx 1.7$ we see that the enhanced mixing increases the minimum initial orbital period required for homogeneous evolution. 

\subsection{Fractions of homogeneously evolving systems \label{sec:fracs}}

The enhancement of mixing in the Enhanced Grid enables systems with lower primary masses to undergo homogeneous evolution, as compared to the Standard Grid. We see from Fig. \ref{fig:gridResults} that the Enhanced Grid systems with $q_i=0.9$ and primary masses as low as log$_{10}(M_1/M_{\odot}) =1.5$ are able to form double helium stars, while in the Standard Grid this limit is log$_{10}(M_1/M_{\odot}) =1.7$. Similarly for systems with $q_i=0.7$, the minimum initial primary mass that evolves homogeneously is shifted from log$_{10}(M_1/M_{\odot}) =1.8$ to 1.6. These relatively low mass double helium stars will have a large effect on predicted numbers of such systems owing to the initial mass function strongly favouring lower primary masses.

The predicted fractions of binary systems that will evolve to form double helium stars can be estimated by giving a weight to each model using a Salpeter distribution of masses for Star 1 ($dN/dm \propto m^{-2.35}$) and a flat distribution in log$_{10}(P)$. The weight of each model is given by 

\begin{align}
W(M_i, P_i) = \frac{\int _{log_{10}(P_i - \Delta P)} ^ {log_{10}(P_i + \Delta P)} dlog_{10}(P)\int _{M_i - \Delta M} ^ {M_i + \Delta M} m^{-2.35} dm}{\int _{log_{10}(0.5)} ^ {log_{10}(2.0)} dlog_{10}(P)\int _{10^{1.4}} ^ { 10^{2.0}} m^{-2.35} dm},
\end{align} 
where $\Delta P$ is the half difference between the period spacing of the models (0.05 days in this work), and $\Delta M$ is the corresponding half difference between the mass spacing of the models in log space, such that $M_i + \Delta M = 10^{(log(M/M_{\odot}) + 0.05)}$

Summing the weights of models that formed double helium stars gives an estimate of the fraction of systems forming a double helium star with initial period spanning 0.5-2.0 days, Star 1 mass spanning log$(M/M_{\odot})=$ 1.4-2.0 at each initial mass ratio. These values are tabulated in Table \ref{tab:homofrac}. Finally by assuming a flat initial mass-ratio distribution, integrating over $q_i$ gives the fraction of binary systems predicted to form double helium stars in our parameter space. We calculate that the models with enhanced mixing produce 2.4 times more double helium stars than those without.

{\renewcommand{\arraystretch}{1.2}
\begin{table*}[h]
\caption{Fractions of binary systems in our model grids that are predicted to form double helium stars for each initial mass-ratio. A Salpeter primary mass distribution ($dN/dM \propto M^{-2.35}$), a flat distribution in log of initial orbital period and a flat initial mass-ratio distribution have been assumed. The total is given by the integral over mass-ratio.}
\centering
\begin{tabular}{llllll}
\hline
\hline
{$q_i$} & 1.0                   & 0.9                   & 0.7                   & 0.5 & Total                \\ \hline
Standard Grid   & 5.81$\times 10^{-3}$ & 2.92$\times 10^{-3}$ & 1.15$\times 10^{-3}$ & 0   & 9.58$\times 10^{-4}$ \\ 
Enhanced Grid   & 8.73$\times 10^{-3}$ & 7.61$\times 10^{-3}$ & 4.14$\times 10^{-3}$ & 0   & 2.28$\times 10^{-3}$ \\ \hline
\end{tabular}
\label{tab:homofrac}
\end{table*}

\subsection{Occurrence of mass-transfer \label{sec:mt}}

As systems with larger orbital periods undergo homogeneous evolution, the separation of the stars can become large enough to avoid one component evolving to fill its Roche-lobe.{ We note that although homogeneously evolving stars do increase their radii slightly during hydrogen burning, such that stars that are detached at the zero-age-main-sequence may evolve into contact, this increase is very small compared to that seen during standard evolution.} 

In our Standard Grid models only a few systems with primary masses near 100\Msol and initial mass-ratio near 1 manage to avoid contact phases. However in the Enhanced Grid, {owing to the larger orbital periods for homogeneously evolving systems,} many systems are able to avoid binary interaction. For the Enhanced Grid models with $q_i=1.0$, approximately half of the systems forming double helium stars do so without any contact phases occurring. While the number of systems avoiding interaction decreases as the initial mass-ratio becomes more extreme, such systems are remarkable because they produce double helium stars with mass-ratios close to their initial mass-ratio. In the Enhanced Grid, double helium stars that avoid interaction are produced at initial mass-ratios as extreme as $q_i=0.7$, whereas in the Standard Grid no non-interacting systems exist at $q_i=0.7$ and only two are present at $q_i=0.9$.

\begin{table}[h]
\caption{Of systems that form a double helium star, the fractions of those that do not undergo mass-transfer for each initial mass ratio. A Salpeter primary mass distribution ($dN/dM \propto M^{-2.35}$), a flat distribution in log of initial orbital period and a flat initial mass-ratio distribution have been assumed. The Total is the integrated fraction of non-interacting double helium stars to all double helium stars, with the number in brackets excluding $q_i=1.0$ systems.}
\centering
\begin{tabular}{llllll}
\hline
\hline
{$q_i$} & 1.0                   & 0.9                   & 0.7                   & 0.5 & Total                \\ \hline
Standard Grid   & 0.095 & 0.087 & 0 & 0   & 0.069(0.049) \\ 
Enhanced Grid   & 0.371 & 0.233 & 0.194 & 0   & 0.245(0.213) \\ \hline
\end{tabular}
\label{tab:homofrac2}
\end{table}

Table \ref{tab:homofrac2} gives the proportion of double helium star systems that do not undergo mass-transfer. We see that only 5\% of double helium stars in the Standard Grid avoid mass-transfer episodes, while the for the Enhanced Grid this figure is around 20\%. This means that in the Standard Grid, almost all of the double helium star systems have a mass-ratio of unity, whereas for the Enhanced Grid, some 20\% of double helium stars remain close to their initial mass-ratio. 

To explore the final mass-ratios of our model systems, we plot the mass of each component at core helium exhaustion in Fig. \ref{fig:massRatios}. Systems with $q_i=1$ remain at their initial mass ratio and so are not plotted. If it is assumed that at the end of helium burning, the star collapses directly to a black hole without any mass loss, Fig. \ref{fig:massRatios} shows the predicted mass-ratios of binary black holes formed through the chemically homogeneous channel. Comparing the Standard with the Enhanced Grid, we see that the enhanced mixing enables binary black holes with mass-ratios as low as 0.8 to be produced. Without enhanced mixing the final systems all have mass-ratios very close to unity. It is also seen that the Enhanced Grid produces systems with component masses as low as 30 \Msol , whereas the Standard Grid can only make carbon stars with masses not lower than 40 \Msol.


\begin{figure*} 
	\includegraphics[width=0.95\linewidth]{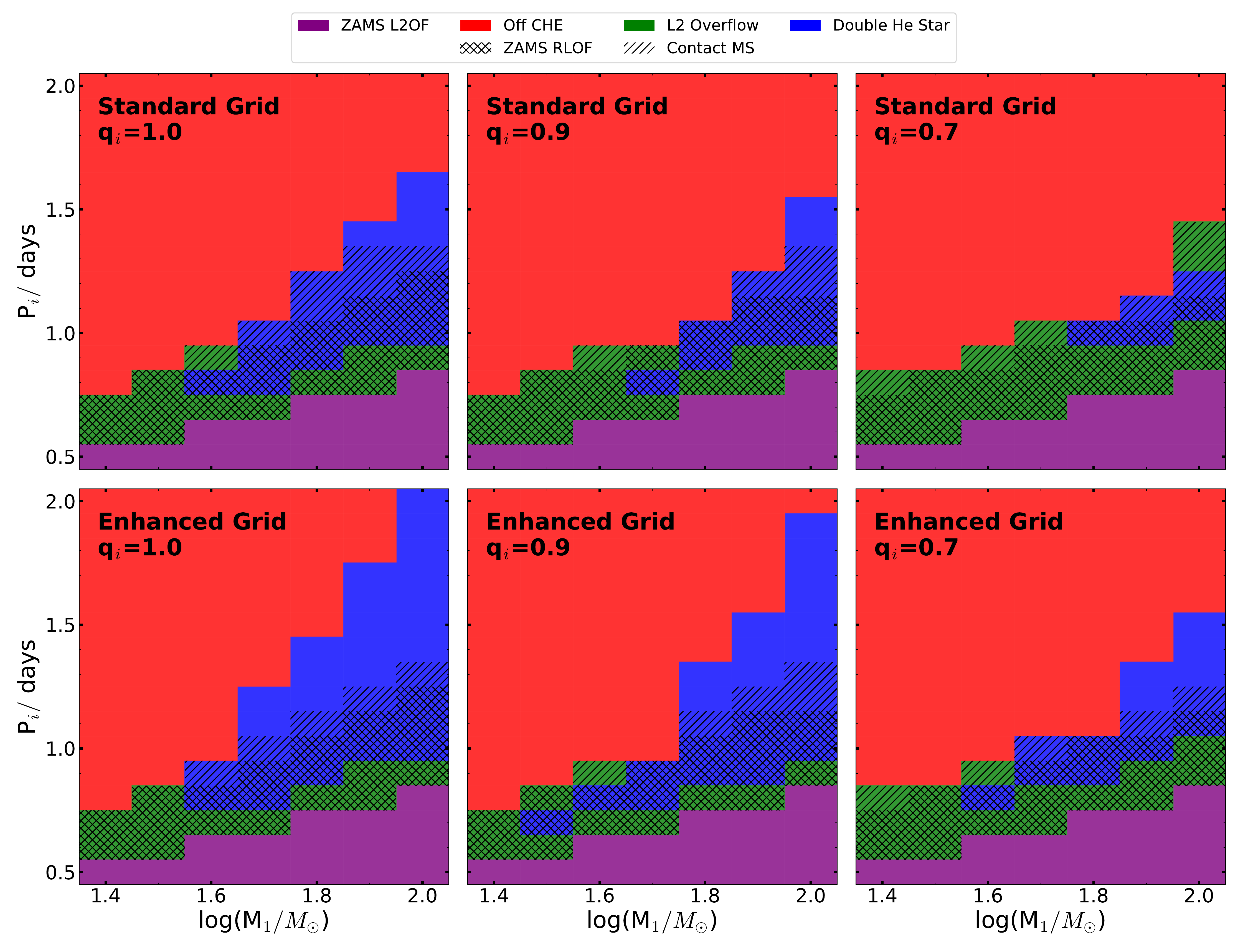}
	\centering
	\caption{Final configurations of models in the Standard Grid (top row) and Enhanced Grid (bottom row) for initial mass ratios $q_i=1.0,0.9,0.7$. The colours indicate where overflow at the second Lagrangian point occurred at the zero-age-main-sequence or during evolution (purple and green respectively), where the difference between central and surface helium mass fraction exceeds 0.2 (red) and where homogeneous evolution occurs resulting in a double helium star (blue). Single hatched cells indicate that the system underwent contact on the main-sequence, and cross-hatched that contact occurred on the zero-age-main-sequence.  }
	\label{fig:gridResults} 
\end{figure*}


\begin{figure}[h!] 
	\includegraphics[width=0.95\linewidth]{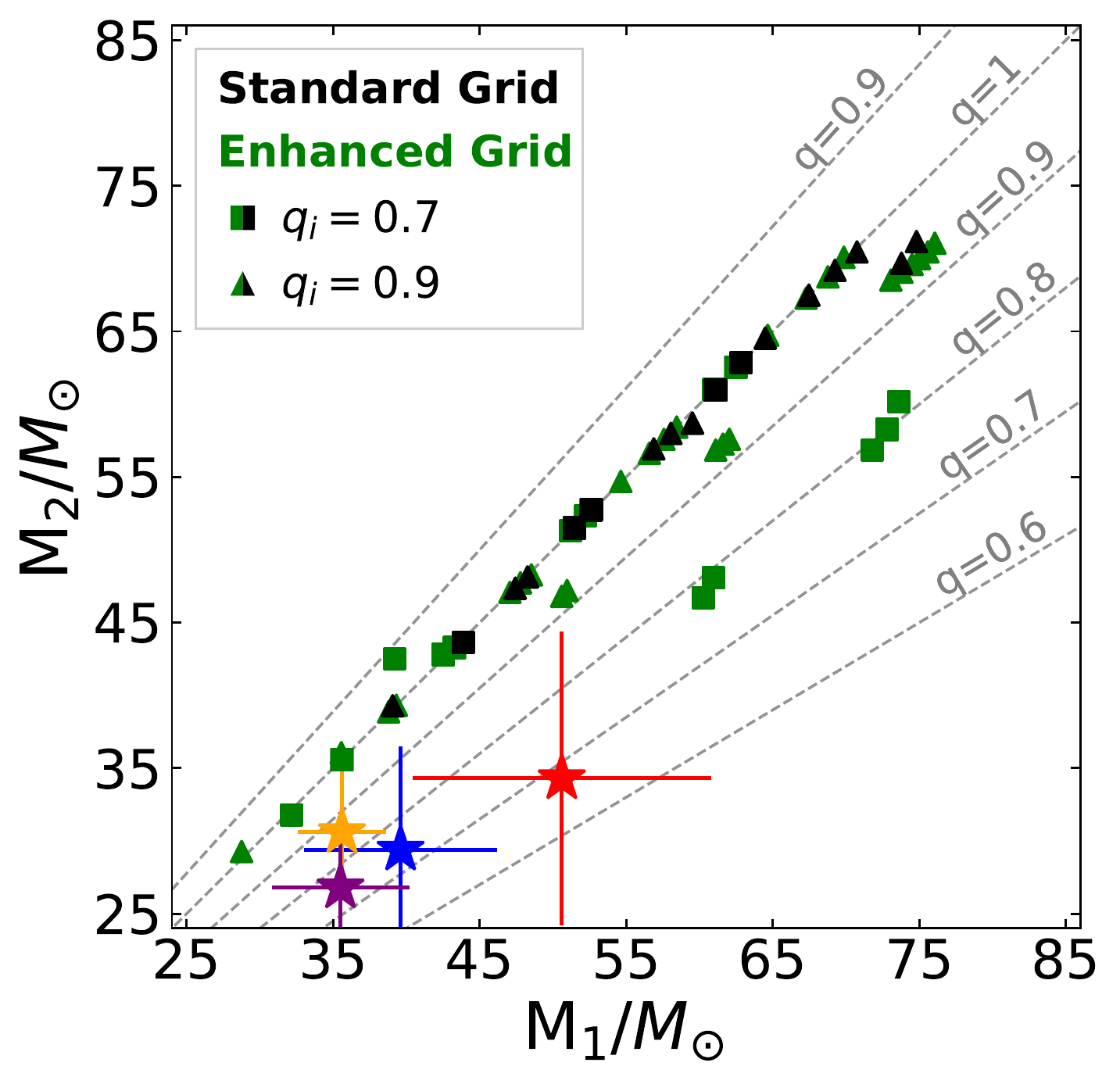}
	\centering
	\caption{ Masses of Star 1 (x-axis) and Star 2(y-axis) at core helium depletion for systems with $q_i=0.9,0.7$ (plotted as triangles and squares respectively). Models in the Standard Grid (without enhanced mixing) are marked in black, those in the Enhanced Grid (with enhanced mixing) in green. For clarity systems with $q_i=1$ are not plotted as they remain at their initial mass ratio. The grey dotted lines show various mass-ratios as marked. Red, blue, yellow and purple stars represent measured progenitor masses of gravitational wave events GW170729, GW170823, GW150914 and GW170818 respectively \citep{2016PhRvX...6d1014A,2018arXiv181112907T}. }
	\label{fig:massRatios} 
\end{figure}

\section{Discussion \label{sec:disc}}
\subsection{Other possible mixing mechanisms \label{sec:othermixing}}

{One should keep in mind that grand-scale circulations are not the only proposed mixing mechanisms operating in binary stars. As our results have shown that binary evolution calculations are sensitive to even a modest increase in mixing efficiency, it of interest to discuss other processes which may have an impact on internal mixing. } 

{The interaction of internal circulation currents with the convective core can enhance mixing at the core-envelope boundary in a way that mimics convective overshooting \citep{2018MNRAS.480.5427J}. }

\citet{2018MNRAS.475.4579V} suggest that certain magnetically induced instabilities can arise in the radiative zones of binary stars, causing mixing and the induction of significant magnetic fields. Such a mechanism may explain the observed magnetism of stars such as Vega. Furthermore, massive stars in close proximity to each other can be imagined to interact magnetically, which may further effect or result in mixing processes. Indeed magnetic fields have been observed in both B-type components of the binary system $\epsilon$ Lupi by \citet{2015MNRAS.454L...1S}, who conclude that the two fields are interacting.

Another agent which can mix material inside a star is pulsation. Massive stars can be pulsationally unstable, where pressure and gravity waves travel throughout the star. These waves are expected to cause some degree of mixing \citep{2014IAUS..301..185A,2017ApJ...848L...1R}. What is more, when there exists a resonance between the orbital and pulsational periods of a binary star,  pulsational instabilities can be strongly excited  \citep{1995ApJ...449..294K, 2001A&A...366..840W}. Such behaviour is observed in so-called heartbeat stars \citep{2017MNRAS.472.1538F}.


\subsection{Implications for black hole merger events\label{sec:GWevents}}

Although the models presented here were computed at metallicity Z= \Zsol /50 and the metallicity will have an effect on the final fates of binary systems (through the stellar radii and mass-loss rates), it is worth comparing our models with observed gravitational wave events. The double black-hole systems produced by our models are all expected to have merger timescales much shorter than the Hubble time  \citep{2016A&A...588A..50M}, and so are deemed as viable progenitors for gravitational wave events.

Shown in Fig. \ref{fig:massRatios} are the component masses of the currently detected gravitational wave events with secondary masses greater than 25\Msol \citep{2016PhRvX...6d1014A,2018arXiv181112907T}. Although our model grid is fairly coarse, a comparison between the models and observed black-hole merger events reveals that it would be very difficult for our Standard Grid models to produce double black holes with such low component masses or mass-ratios as low as those observed. {On the other hand, although our Enhanced Grid models do not fit the observations perfectly, taking into account metallicity effects and the coarseness of the model grid, we conclude that the Enhanced Grid models do not rule out the possibility that the four merger events could have been produced by the chemically homogeneous channel. }

The detailed evolutionary models of \citet{2016A&A...588A..50M} investigated black-hole merger events from the chemically homogeneous channel without modelling companion-induced circulation and predicted that below the pair-instability supernova gap, every double black-hole system has a mass-ratio of unity, in agreement with our Standard Grid models. As demonstrated by our Enhanced Grid models, the inclusion of companion-induced circulation changes this picture significantly. {We predict that around 20\% of double black-hole systems will have avoided mass-transfer, and estimate from Tables \ref{tab:homofrac} and \ref{tab:homofrac2} that 12\% of systems have final mass-ratios in the range $0.80< q <0.95$ and 8\% with $0.60< q <0.80$.} Although our models predict the lowest achievable mass-ratio to be approximately 0.8, at lower metallicities (\citet{2020arXiv200211630D} find that systems with metallicities as low as $10^{-5}$\Zsol contribute to observable double black-hole merger events) weaker stellar winds will allow systems to remain much closer to their initial mass-ratios, thus making the production of double black-hole systems with mass-ratios around 0.7 entirely possible.

As our Enhanced Grid models produce {2.4 times the number of double helium stars} than those of the Standard Grid (see Sec. \ref{sec:fracs}), the inclusion of companion-induced circulation would increase the predicted rate of double black-hole merger events by a similar factor. Although we have considered only one metallicity, it may reasonably be assumed that the inclusion of companion-induced circulation will affect other metallicities similarly.  \citet{2020arXiv200211630D} calculate the aLIGO detection rate for systems below the pair-instability supernova gap  to be 245 yr$^{-1}$,{ with our results suggesting that this rate should be revised upwards, by a factor of 2.4, to some 590 events per year, equating to nearly two detections per day. It has been predicted that gravitational wave events from the homogeneous channel could well be the dominant source for future detections \citep{2016A&A...588A..50M,2016MNRAS.458.2634M,2020arXiv200211630D}. The results of this work serve to reinforce such statements. }

\section{Conclusions\label{sec:conc}}

We have considered the mathematical results of \citet{1982ApJ...261..265T,1982ApJS...49..317T} for the steady-state circulation in the radiative zones of a single rotating star and a tidally locked binary star. By making a suitably justified approximation (namely that some components of the velocity field can be ignored), {it has been shown that the root-mean-squared circulation velocity in the radial direction in a binary component may be expressed as a velocity enhancement to the Eddington-Sweet circulation.} This enhancement is a simple function depending only on the masses of stars in the binary system. Such a result allows one to easily include the effects of companion-induced circulation in one-dimensional stellar evolution codes.

We investigated the effects of companion-induced circulation on tidally locked binary systems with masses below the pair-instability supernova gap using the detailed stellar evolution code MESA. We find that at initial mass-ratios in the range 0.7-1.0, binary systems with larger initial periods and slightly lower initial masses are able to form double helium stars through homogeneous evolution, compared to the standard mixing scenario. The circulation velocity enhancement also results in fewer systems undergoing mass-transfer phases. Under the assumptions of a Salpeter primary mass distribution, a distribution of the initial orbital period that is flat in logarithmic space and a flat initial mass-ratio distribution we calculate that in our parameter space models with enhanced mixing produce 2.4 times more double helium stars than models with standard mixing. 

Our findings have strong implications for the formation of merging double black-hole systems. Previous detailed modelling has shown that the homogeneous evolution channel is only able to form double black-hole binaries with mass-ratios very close to one \citep{2016A&A...588A..50M}. The inclusion of companion-induced circulation into the models alters this, as we find double black-hole systems with mass-ratios as low as 0.7 $\sim$ 0.8 are able to be produced from tidally locked binary stars. Our models show that systems with uneven mass-ratios are expected to make up some 20\% of the total population of merger events. 

Furthermore the previously calculated gravitational wave event detection rates from the homogeneous evolution channel \citep{2016A&A...588A..50M,2016MNRAS.458.2634M,2020arXiv200211630D} should be revised upwards in light of our findings, by a factor of 2.4. This would lead to a predicted aLIGO detection rate of nearly 600 events per year, making the homogeneous evolution channel a certain contender for the dominant production source of double black-hole merger events.

\bigbreak 
Acknowledgements: {We thank our anonymous referee for useful comments on an earlier version of this manuscript.} GK acknowledges support from CONACYT 252499 and UNAM/PAPIIT IN103619

\addcontentsline{toc}{section}{References}
\bibliography{bib}
\bibliographystyle{aa_url}

\begin{appendix}

\onecolumn
\section{Calculation of the rms circulation velocity \label{sec:case2calc}}
{Here we give details of the calculation of the rms circulation velocity for our extreme case considered in Sec. \ref{sec:Ex2}, where all $U_k$ terms contribute equally to the circulation velocity.}
In the limit that $U_2 \approx U_3 \approx U_4$, Eq.\ref{Eq:binaryR} becomes 
\begin{align}
U_{r}(r, \nu, \mu)&= \epsilon U_S(r) [ Q\sum_{k=2}^{4}   P_k(\nu) +  P_2(\mu)].
\end{align}
and we have
\begin{align}
& {\int_{\varphi =0}^{\varphi=2\pi} \int_{\theta =0}^{\theta=\pi} v_{r}^2 d\theta d\varphi } =
 \epsilon^2 U_S(r)^2 \int_{\varphi =0}^{\varphi=2\pi} \int_{\theta =0}^{\theta=\pi} \Bigg[Q^2 \bigg(    \sum_{k=2}^{4}   P_k(cos(\varphi)sin(\theta))\bigg)^2  + P_2(cos(\theta))^2 
+ 2QP_2(cos(\theta))\sum_{k=2}^{4}   P_k(cos(\varphi)sin(\theta)) \Bigg]d\theta d\varphi. 
\end{align}
For convenience we shall let 
\begin{align}
{\int_{\varphi =0}^{\varphi=2\pi} \int_{\theta =0}^{\theta=\pi} v_{r}^2 d\theta d\varphi } =&  \epsilon^2 U_S(r)^2  \bigg[A+ B +C\bigg] .
\end{align}
Where $A,B,C$ are defined below. 
\begin{align}
{A }   &=Q^2 \int_{\varphi =0}^{\varphi=2\pi} \int_{\theta =0}^{\theta=\pi} \Bigg[ P_2(\nu)^2 + P_3(\nu)^2 + P_4(\nu)^2 + 2P_2(\nu)P_4(\nu) \Bigg]d\theta d\varphi ,
\end{align}
which has the analytic solution
\begin{align}
{A }   &= Q^2 \pi^2 \frac{682039260864257}{1000000000000000} \approx 0.68 Q^2 \pi^2.
\end{align}
Now moving onto $B$,   
\begin{align}
B &= \int_{\varphi =0}^{\varphi=2\pi} \int_{\theta =0}^{\theta=\pi}  P_2(\mu)^2 d\theta d\varphi ,
\end{align}
giving simply
\begin{align}
B &=  \frac{11\pi^2}{16} = 0.6875 \pi^2 .
\end{align}
Lastly we tackle $C$, 
\begin{align}
C  &=2Q \int_{\varphi =0}^{\varphi=2\pi} \int_{\theta =0}^{\theta=\pi}  P_2(\mu)P_2(\nu) + P_2(\mu)P_3(\nu) + P_2(\mu)P_4(\nu)d\theta d\varphi,
\end{align}
giving 
\begin{align}
C  &= -2Q\pi^2 \frac{235}{1024} \approx -Q\pi^2 \times 0.4590 .
\end{align}
Combining $A,B,C$ gives 
\begin{align} 
\vrms & \approx \epsilon |u_S(r)|  \sqrt{0.3410 Q^2  +0.34375  -  0.2295 Q}  \, . 
\end{align}

\section{An investigation of the functions $U_k$ \label{sec:Ukcomparison}}

We will use simple models to investigate how the inviscid radial circulation velocity magnitudes arising from a binary companion, $U_2, U_3, U_4$ are related to each other. Specifically we aim to see which of our extreme cases considered in Section \ref{binary_comp}, $ U_2 >> U_3 >> U_4$ or $ U_2 \approx U_3 \approx U_4$ is most realistic. To accomplish this we employ the theory of polytropes, for a detailed description of polytropes one should refer to \citet{1957isss.book.....C}. A polytrope is an object which obeys the following relation between pressure and mass density 
\begin{align}
P = K_P \rho ^{\frac{n+1}{n}} \label{Eq:polytrope}
\end{align}
with $K_P$ being a constant and $n$ the polytropic index. It can be shown \citep{1957isss.book.....C} that the polytropic constant, $K_P$, is given by 
\begin{align}
K_P= \frac{G}{n+1} \Bigg[ \frac{4 \pi}{\xi_1 ^{n+1}} \bigg(-{\frac{d\theta}{d\xi} }\rvert_{\xi_1}\bigg)^{1-n} \Bigg] ^{\frac{1}{n}} M^{\frac{n-1}{n}} R^{\frac{3-n}{n}} , \label{Eq:Kp}
\end{align} 
where $\theta = (\rho / \rho_c)^{\frac{1}{n}}$ is the solution to the Lane-Emden equation and $\xi$ is the radial co-ordinate divided by normalisation factor $\alpha$, $\xi = r / \alpha$ . The radius of the star is represented by $\xi_1$ , $\xi_1 = R / \alpha$ and is found by setting $\theta =0$. 

{By combining the equation of hydrostatic equilibrium, 
\begin{align}
p'&= \frac{-Gm \rho}{r^2} 
\end{align}
with the relation from \citet{1982ApJS...49..317T} ,
\begin{align}
\rho' &= -\frac{n}{n+1} \frac{-Gm \rho^2}{p r^2},
\end{align}
one can show that 
\begin{align}
\frac{\rho'}{p'} &=  \frac{n}{n+1} \frac{\rho}{p}. 
\end{align}
}
Thus
\begin{align}
 \frac{\rho' \rho}{p'} &= \frac{n}{n+1} \frac{\rho ^2}{p}. \label{Eq:term}
\end{align}
The fraction $\frac{\rho ^2}{p}$ can be rewritten using the polytropic relation Eq. \ref{Eq:polytrope} so that 
\begin{align}
\frac{\rho' \rho}{p'} &= \frac{n}{n+1} \rho^{\frac{n-1}{n}} K_P^{-1} .\label{Eq:A7}
\end{align}

Now we will make use of the fact that
\begin{align}
\rho_c = \frac{M}{4 \pi R^3} \xi_1 \Bigg[- \frac{d\theta}{d\xi} |_{\xi_1} \Bigg] ^{-1}, \label{Eq:polyrhoc}
\end{align}
  (see \citet{1957isss.book.....C}) to rewrite Eq. \ref{Eq:A7} in terms of $\frac{\rho}{\rho_c} = \theta ^n$ . This gives
\begin{align}
\frac{\rho' \rho}{p'} &= \frac{n}{n+1} \Bigg[ \frac{M \xi_1}{4\pi R^3} \Bigg]^{\frac{n-1}{n}} \Bigg[- \frac{d\theta}{d\xi} |_{\xi_1} \Bigg] ^{\frac{1-n}{n}}  K_P^{-1} \theta ^{{n-1}} .
\end{align}
Lastly we use the form of the polytropic constant, $K_P$ given in Eq. \ref{Eq:Kp} to produce
\begin{align}
\frac{\rho' \rho}{p'} &= \frac{n }{4 \pi G} \xi_1 ^{2}   R^{ -2}  \theta ^{{n-1}} . \label{Eq:termsimple}
\end{align}

To determine $U_k(r)$, we first need to solve the differential equation Eq. \ref{Eq:phik}, which when combined with Eq. \ref{Eq:termsimple} becomes 
\begin{align}
\phi_k '' + \frac{2}{r} \phi_k' + \phi_k \bigg[ n \zeta_1 ^{2}   R^{ -2}  \theta ^{{n-1}} - \frac{k(k+1)}{r^2}\bigg] =0 . \label{Eq:DE_poly}
\end{align}

We must now convert this from a differential equation in $r$ to one in $\xi$ by defining $F_k(\xi)$ as an equivalent of $\phi_k(r)$, thus 
\begin{align}
\frac{d \phi_k}{dr} = \frac{d F_k}{d \xi} \frac{d \xi}{dr} = \frac{d F_k}{d \xi} \frac{1}{\alpha} .
\end{align}
We may now recast Eq. \ref{Eq:DE_poly} as 
\begin{align}
\frac{d^2 F_k}{d \xi^2}  + \frac{2}{\xi } \frac{d F_k}{d \xi}  + F_k \bigg[n    \theta ^{{n-1}} - \frac{k(k+1)}{\xi^2 }\bigg] &=0 . \label{Eq:finalDE}
\end{align}
Happily this differential equation does not depend on the model's mass or radius, only the polytropic index, as $\theta$ is a function of $n$.

One last step is to redetermine the constant $c_k$ in terms of $\xi$ and $F_k(\xi)$ so that it becomes 
\begin{align}
c_k &= - \omega_0^2 \frac{M_{comp}}{M+M_{comp}} \bigg(\frac{R}{d}\bigg)^{(k-2)} \frac{(2k+1)(\xi_1 \alpha)^2}{(k+1)F_k( \xi_1) + \xi_1  F'_k( \xi_1)} . \label{Eq:c_final}
\end{align}

Eq. \ref{Eq:finalDE} must be solved numerically, not least because in general $\theta$ has no analytic solution. We shall choose to solve for an $n=3$ polytrope, giving $\xi_1= 6.8968$ and $- \frac{d\theta}{d\xi}=0.0424 $. This polytrope describes the Eddington Standard Model, where energy transport is partially radiative, so is the best choice for the envelope of a massive star. In terms of the Emden variables the functions $U_k$ defined in Eq \ref{Eq:USbinary} become 
\begin{align}
U_k(r) &= \frac{2L\xi^4 \alpha^4}{G^2 m^3} \frac{n+1}{n-1.5}c_k \left[\frac{dF_k}{d\xi}\frac{1}{\alpha} + (\frac{2}{\xi \alpha}- \frac{3}{\xi \alpha})F_k\right] .\label{Eq:finalDe}
\end{align}
The mass, $m$, is expressed as an integral of the density
\begin{align}
m(\xi) = 4 \pi \alpha^3 \rho_c \int _0^{\xi} \xi'^2  \theta^n d\xi' ,
\end{align}
which when combined with Eq. \ref{Eq:polyrhoc} leads to 
\begin{align}
m(\xi) =  \alpha^3 \xi_1 \frac{M}{R^3} \Bigg[-\xi_1^2 \frac{d\theta}{d\xi} |_{\xi_1} \Bigg] ^{-1} \int _0^{\xi} \xi'^2  \theta^n d\xi' . \label{Eq:polytropem}
\end{align}

Eqs \ref{Eq:finalDe} and \ref{Eq:polytropem} can be solved numerically to provide the functions $U_k(r)$. These functions are plotted in Fig. \ref{fig:Ukcomp} for an $n=3$ polytrope with $M=$ 3\Msol, $R=1.75$\Rsol, $L=$ 93 \Lsol and a companion mass of  $M_{\textrm{comp}}=$ 3\Msol. The luminosity is assumed to be concentrated in a central point. From Eq. \ref{Eq:c_final} we see that for small values of $R/d$, the constants $c_k$ decrease sharply with increasing $k$. This means that the functions $U_k(r)$ will be most similar to each other for high $R/d$ values, therefore we have chosen the ratio of stellar radius to orbital separation to be 0.35, which represents a system that is very close to contact (for mass-ratios of 1 a contact system has  $R/d=0.379$ \citep{1983ApJ...268..368E}).

\begin{figure} 
	\includegraphics[width=0.55\linewidth]{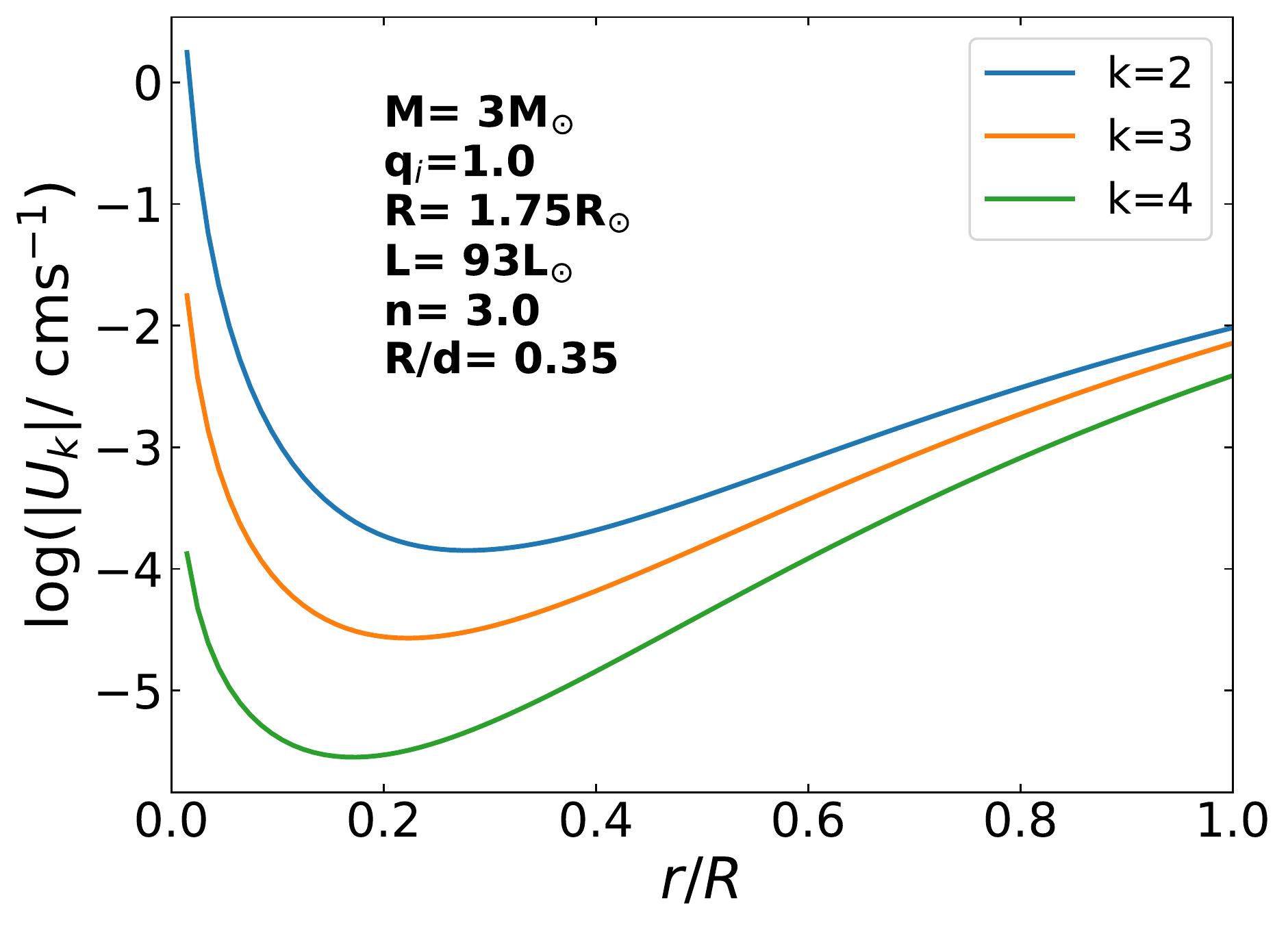}
	\centering
	\caption{The inviscid radial circulation velocity magnitudes, $U_k(r)$ plotted as a function of normalised radius for an $n=3$ polytrope with $M=$ 3\Msol, $R=1.75$\Rsol, $L=$ 93 \Lsol and a companion mass of  $M=$ 3\Msol. The ratio of stellar radius to orbital separation is set to be 0.35, representing a near-contact system. }
	\label{fig:Ukcomp} 
\end{figure}

A more interesting quantity to consider is the ratios of $U_3$ and $U_4$ with $U_2$. We note that such ratios will not depend on the specifics of our chosen star (mass, luminosity and radius) because the solutions to Eq. \ref{Eq:finalDE} only depend on the $k$ value and polytropic index. The only other parameter not divided out is the ratio of stellar radius to orbital separation. Fig.  \ref{fig:Ukcomp2} shows the ratios of the $U_k$ functions for varying $R/d$ values. It is seen that at most $U_3$ is 70\% of $U_2$ and $U_4$ is 40\% of $U_2$. For smaller values of $R/d$, corresponding to detached systems, $U_2$ becomes even greater compared to $U_3$ and $U_4$. Therefore it is concluded that the approximation $U_2 >> U_3 >> U_4$ is more suitable than $U_2 \approx U_3 \approx U_4$.




\begin{figure*} 
	\includegraphics[width=0.9\linewidth]{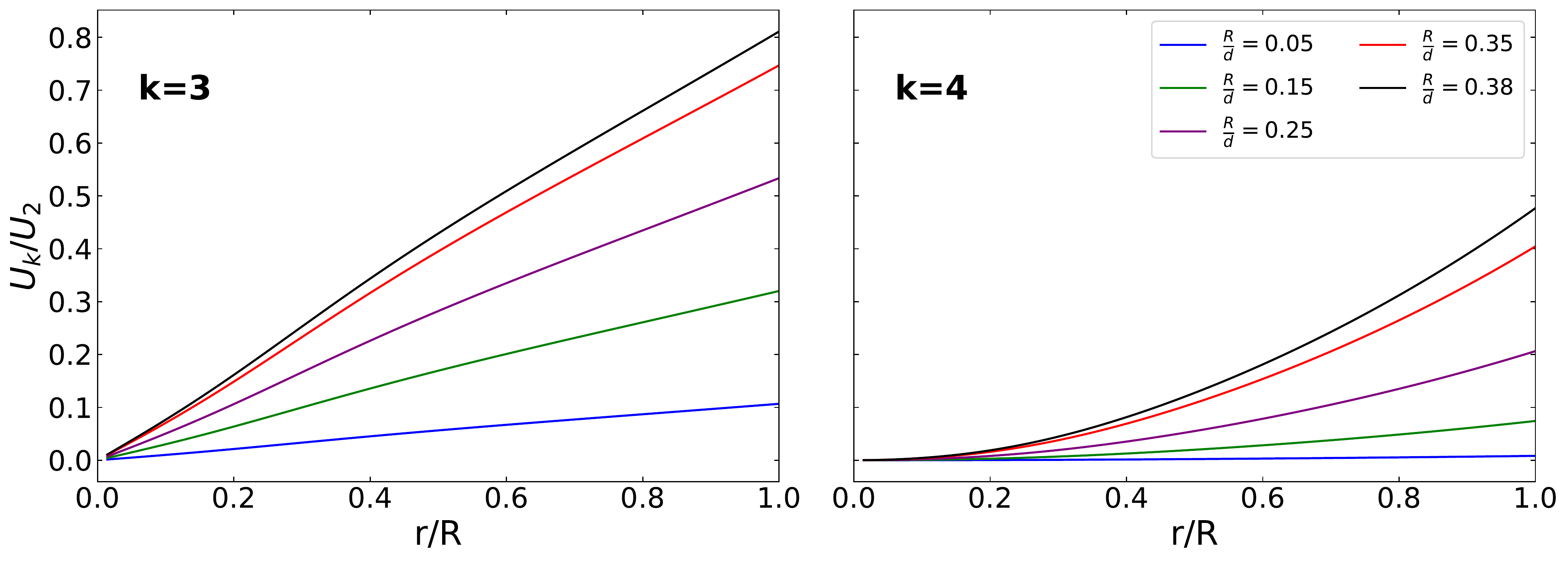}
	\centering
	\caption{ The functions $U_3$ (left panel) and $U_4$ (right panel) as fractions of $U_2$ plotted as a function of normalised radius for the polytropic model in Fig.\ref{fig:Ukcomp}. Various values of the stellar radius to orbital separation, $R/d$ are plotted as coloured lines as given by the legend. $R/d=0.38$ represents a contact system and is hence an upper limit.}
	\label{fig:Ukcomp2} 
\end{figure*}






\end{appendix}

\end{document}